\documentclass{aa}  
\makeatletter
\renewcommand*\aa@pageof{, page \thepage{} of \pageref*{LastPage}}
\makeatother
\usepackage{graphicx}
\usepackage{txfonts}
\usepackage{multirow}
\usepackage{subcaption}
\usepackage[switch]{lineno}
\usepackage{hyperref}
\usepackage{cleveref}
\hypersetup{
    colorlinks=true,
    linkcolor=blue,
    filecolor=magenta,      
    urlcolor=cyan,
    citecolor=blue,
}
\begin{document}

   \title{X-ray spectral fitting with Monte Carlo Dropout Neural Networks}


   \author{A. Tutone
          \inst{1}\fnmsep\inst{2}\fnmsep\thanks{\email{antonio.tutone@inaf.it}},
          A. Anitra\inst{3}\fnmsep\inst{4}, 
          E. Ambrosi\inst{1},
          R. La Placa\inst{5}
          A. D'Aì\inst{1},
          C. Pinto\inst{1},
          M. Del Santo\inst{1},
          F. Pintore\inst{1},
          A. Pagliaro\inst{1}\fnmsep\inst{2}\fnmsep\inst{6},
          A. Anzalone\inst{1}\fnmsep\inst{2}\fnmsep\inst{6},
          T. Di Salvo\inst{4},
          R. Iaria\inst{4},
          L. Burderi\inst{1, 3},
          A. Sanna\inst{3}
          }
   \authorrunning{A. Tutone et al.} 

   \institute{Istituto Nazionale di Astrofisica INAF IASF Palermo, 
                Via Ugo La Malfa 153, Palermo, 
                I-90146, Italy
         \and
             ICSC – Centro Nazionale di Ricerca in HPC, Big Data e Quantum Computing, Italy
         \and Dipartimento di Fisica, Universit\`a degli Studi di Cagliari, SP
Monserrato-Sestu, KM 0.7, Monserrato, 09042 Italy
         \and Dipartimento di Fisica e Chimica - Emilio Segrè,
 Universit\`a di Palermo, via Archirafi 36 - 90123 Palermo, Italy
         \and INAF-Osservatorio Astronomico di Roma, Via Frascati 33, I-00076 Monte Porzio Catone, (RM), Italy
         \and Istituto Nazionale di Fisica Nucleare Sezione di Catania, 
         Via Santa Sofia 64, 95123 Catania, Italy\\
             }

   \date{Received ; accepted }

 
  \abstract
    {The analysis of X-ray spectra often encounters challenges due to the tendency of frequentist approaches to be trapped in local minima, affecting the accuracy of spectral parameter estimation. Bayesian methods offer a solution to this issue, even though computational time significantly increases, limiting their scalability. In this context, neural networks have emerged as a powerful tool to efficiently address these challenges, providing a balance between accuracy and computational efficiency.}
   {This work aims to explore the potential of neural networks to recover model parameters and quantify their uncertainties, benchmarking their accuracy and computational time performance against traditional X-ray spectral fitting methods based on frequentist and Bayesian approaches. This study serves as a proof of concept for data analysis of future astronomical missions, producing extensive datasets that could benefit from the proposed methodology.}
   {We apply Monte Carlo Dropout to a range of neural network architectures to analyze X-ray spectra. Our networks are trained on simulated spectra derived from a multiparameter source emission model, convolved with an instrument response. This allows them to learn the relationship between the spectra and their corresponding parameters while generating posterior distributions. The model parameters are drawn from a predefined prior distribution. To illustrate the method, we used data simulated with the response matrix of the X-ray instrument NICER. We focus on simple X-ray emission models with up to five spectral parameters for this proof of concept.}
   {Our approach delivers well-defined posterior distributions, comparable to those produced by Bayesian inference analysis, while achieving an accuracy similar to traditional spectral fitting. It is significantly less prone to falling into local minima, thus reducing the risk of selecting parameter outliers. Moreover, this method substantially improves computational speed compared to other Bayesian approaches, with computational time reduced by roughly an order of magnitude.
   }
    {Our method offers a robust alternative to the traditional spectral fitting procedures. Despite some remaining challenges, this approach can potentially be a valuable tool in X-ray spectral analysis, providing fast, reliable, and interpretable results with reduced risk of convergence to local minima, effectively scaling with data volume.}
   \keywords{ Methods: data analysis, statistical -- X-rays 
               }

   \maketitle
%

\section{Introduction}

The analysis of X-ray spectra has been essential to reveal the physical characteristics of astrophysical objects like active galactic nuclei (AGN), accreting neutron stars, and black hole binaries. Traditionally, spectral fitting, in which physically motivated model parameters are adjusted to match observed data, has been the primary method used in this domain. In recent years, approaches based on both frequentist and Bayesian statistics have proven to be effective, allowing researchers to infer best-fitting model parameters through the use of software such as \texttt{XSPEC}~\citep{Arnaud_1996}, \texttt{SPEX}~\citep{Kaastra_1996} and \texttt{BXA}~\citep{Buchner_2014}. However, these tools can be computationally demanding, especially as astronomical datasets expand with current missions like XRISM and eROSITA~\citep{Tashiro_2018, Predehl_2021} and future missions such as NewAthena~\citep{Barret_2023}, where both the volume of data and the increase in features from higher spectral resolution will significantly impact data analysis.

Recent advances in machine learning, artificial intelligence that enables computers to learn from data and improve their performance on tasks without being explicitly programmed, have introduced new approaches to X-ray spectral analysis. For example, \citet{Tzavellas_2024} and~\citet{Begue_2024} trained Neural Networks (NNs) to generate spectra from physical parameters, enabling rapid model simulations. Another approach uses NNs to estimate model parameters directly from the observed spectra, bypassing the need for traditional iterative fitting~\citep{Ichinohe_2018, Parker_2022, Barret_2024}. Both methods significantly improve speed and efficiency depending on whether the goal is to generate spectra or extract parameters. However, a major challenge remains in obtaining robust uncertainty quantification, which is crucial for reliable astrophysical inferences. Additionally, many existing methods depend on restrictive preprocessing steps, such as narrowing parameter priors or applying dimensionality reduction techniques. While these steps can improve performance, they often limit the flexibility and generalizability of the models, reducing their applicability to diverse datasets.

Our study builds upon these recent advances by applying Monte Carlo Dropout \citep[MCD;][]{Gal_2015}, a technique to approximate Bayesian inference through NNs, to X-ray spectral fitting. This approach provides a robust framework for parameter estimation, including uncertainty quantification, comparable to traditional Bayesian methods, but with improved computational efficiency. We train NNs using simulated spectra generated from multiparameter emission models, convolved with the instrument response of the NICER X-ray observatory \citep{Gendreau_2012}, which operates in the $0.2 - 12$ keV energy range. The models used to simulate the training dataset are relatively simple, involving up to five parameters, an approach commonly employed in recent studies \citep{Parker_2022, Barret_2024}.

The goal of this work is to demonstrate that NNs, combined with MCD, can reliably recover model parameters and their uncertainties from X-ray spectra. Our method is benchmarked against standard spectral fitting techniques and shows significant improvements in computational speed, allowing for real-time data analysis. By doing so, we lay the groundwork for applying these methods to future astronomical missions, such as NewAthena, which will produce vast and complex datasets that traditional methods may struggle to handle efficiently. Furthermore, this approach proves particularly useful for survey analysis, where the ability to process large volumes of data quickly and accurately is essential. The use of NNs in spectral analysis offers speed and a reduction in the risk of local minima trapping, a common issue in conventional fitting algorithms.

In Section~\ref{sec:methods}, we present the methods used to generate the training datasets (\ref{sec:dataset_generation}), construct the neural network (\ref{sec:network}), and evaluate the model performance on simulated data (\ref{sec:performances}). In Section~\ref{sec:real_data}, we apply our approach to real observational data, demonstrating its effectiveness in practical scenarios. Finally, in Section~\ref{sec:discussion}, we discuss the implications of our findings, highlight the advantages and limitations of our method, and propose potential avenues for future work.


\section{Methods}\label{sec:methods}
The fundamental steps of our methodology are as follows: 
\begin{enumerate}
    \item We begin by generating an extensive dataset of simulated X-ray spectra;
    \item We then train the neural network using the generated datasets;
    \item Finally, we evaluate the model’s performance.
\end{enumerate}
Detailed explanations of each step are provided in the next sections.

\subsection{Dataset generation}\label{sec:dataset_generation}
We used \texttt{PYXSPEC}~\citep{Gordon_2021}, the \texttt{PYTHON} wrapper for \texttt{XSPEC}, to simulate the dataset of spectra for the NN training.
We computed the dataset following ~\cite{Barret_2024}, whose recent and detailed work we adopted as a benchmark to facilitate comparison among different neural network-based methods. By aligning with their approach, we ensure consistency in dataset structure and simulation strategies, enabling a more direct evaluation of the performance of various neural network models.

Initially, we adopted a simple model (Model A) consisting of a {\tt powerlaw} continuum emission multiplied by the Tübingen-Boulder model for the absorption by the interstellar medium (ISM), {\tt tbabs} in \texttt{XSPEC}. We set the most updated ISM abundances and cross-sections \citep{Wilms_2000, Verner_1996}.
For the simulations, we used the NICER response file {\tt ni1050300108mpu7}, corresponding to observation {\tt 1050300108} in the NICER archive (see Sec.~\ref{sec:real_data}).
The simulated spectra were rebinned with 5 consecutive channels per bin, resulting in $\sim140$ bins per spectrum, covering the energy range from $0.3$ to $10$ keV. The range of the simulated parameters, for the Model A configuration, is specified in Table~\ref{Tab:model_A}. The parameters were randomly sampled within the specified ranges, avoiding using a grid search approach. We assumed uniform distributions in linear space for the interstellar equivalent absorption column, $N_H$, the Photon Index of the power law component, and a logarithmic distribution for its normalization. We used the \texttt{fakeit} command in \texttt{XSPEC} to generate the synthetic spectra. 

Following the approach of \citet{Barret_2024}, we began by selecting a set of parameter values from Model A to define a reference model: $N_H = 0.2 \times 10^{22}$ cm$^{-2}$, Photon Index = $1.7$, and NormPL = $1$ photons/(keV cm$^2$ s) at $1$ keV. Using these parameters, we simulated two reference spectra with different exposure times: $1$ second and $10$ seconds, corresponding to approximately $2 \times 10^3$ and $2 \times 10^4$ total counts, respectively.

Using these reference spectra as a baseline, we generated two distinct datasets. We simulated $10^4$ spectra for each exposure time by varying the model parameters within the ranges specified in Table~\ref{Tab:model_A} while keeping the exposure time fixed to $1$ second or $10$ seconds. These datasets are referred to as the "Dataset-short" (for 1-second exposure) and the "Dataset-long" (for 10-second exposure). This approach captures realistic spectral variability typical of different classes of X-ray emitting sources while also reflecting the distinct signal-to-noise regimes associated with different exposure times. 

\begin{figure}[t]
    \centering
    \includegraphics[width=1\linewidth]{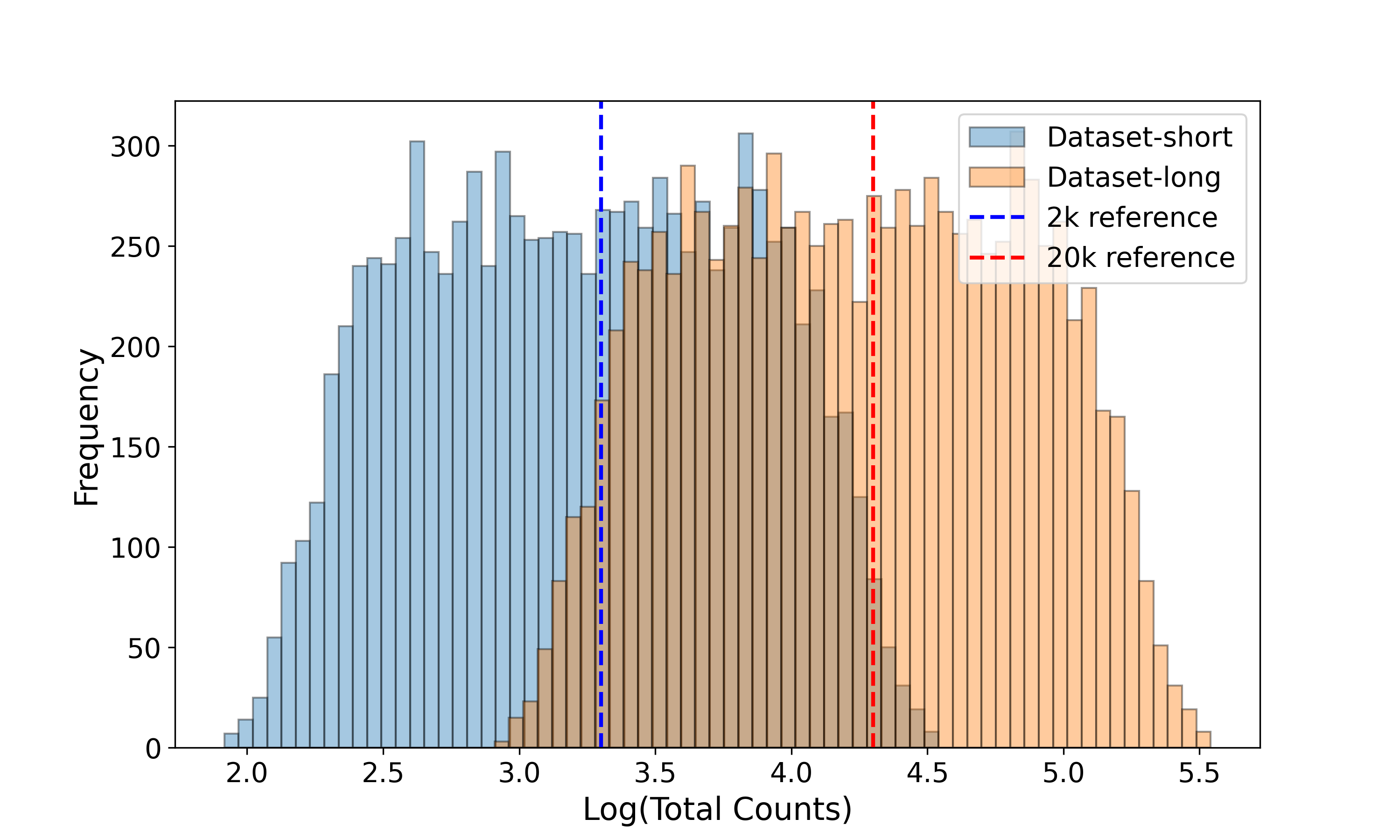}
    \caption{Distribution of total counts for the spectra in the two datasets. The blue histogram represents the dataset based on the reference spectrum with approximately $2 \times 10^3$ total counts (the Dataset-short). In comparison, the orange histogram corresponds to the dataset based on the reference spectrum with approximately $2 \times 10^4$ total counts (the Dataset-long). Both distributions illustrate the variability in total counts due to the range of parameters used in the simulations. Vertical dashed lines show the position of the two reference spectra.}
    \label{fig:histograms}
\end{figure}

In Fig.~\ref{fig:histograms}, we present the distribution of the total counts for the spectra in our two datasets, the Dataset-short and the Dataset-long, respectively. Unlike in~\citet{Barret_2024}, our datasets show some overlap in total counts. This difference arises because~\citet{Barret_2024} initially narrows the parameter range around the reference spectrum. Specifically, they use a ResNet classifier~\citep{He_2016} to quickly approximate the values of the parameters, effectively reducing the parameter search space before the main analysis. Alternatively, they performed a coarse inference, in which the network is trained with limited samples. The posterior conditioned on the reference observation is then used as a restricted prior for subsequent analysis. While these methods are powerful and effective, as demonstrated in their work, our focus is on evaluating the predictive power of our neural network and its ability to estimate uncertainties. By avoiding parameter range narrowing techniques, we aim to maintain the generality of our model after training, making it more versatile for a broader range of scenarios.
 As in~\citet{Barret_2024}, we opted not to include any instrumental background in our simulations. Nonetheless, if an analytical model for the background were available, the network could be trained to account for both the source and background spectra by expanding the number of model parameters. This approach, however, would necessitate a larger training dataset, which could, in turn, lead to longer inference times.

As is common practice in neural network applications~\citep{Bishop_1995, Goodfellow_2016}, both the input spectra and the output parameters are preprocessed before being fed into the network to improve the effectiveness of the training process. Our tests showed that input normalization plays a fundamental role in stabilizing the optimization process, while the parameter normalization mainly helps in balancing the predictive accuracy across different parameters. We used the \texttt{scikit-learn} package~\citep{Pedregosa_2011} for this task. Specifically, we standardized the input spectra by applying the \texttt{StandardScaler} to each energy bin independently, adjusting the values so that each bin across all spectra has a mean of zero and a variance of one. This step was necessary to prevent high-count regions from dominating the optimization and to ensure that all spectral features contribute effectively to the learning process. Additionally, by bringing all spectra onto a similar scale, this normalization helps preventing numerical instabilities that could cause the training process to diverge rather than to converge.

We applied normalization for the output parameters, which rescales the values to fit within the $[0, 1]$ range using the \texttt{MinMaxScaler}. Our tests indicated that this step helped ensuring that all parameters were treated with comparable importance by the network. Without this rescale, parameters with large numerical values tended to have a stronger influence on the optimization, potentially leading to imbalances in predictive accuracy. Such an approach improved the consistency of the model’s predictions by bringing all parameters onto the same scale.

Each dataset is then divided into three subsets: $60\%$ of the data is used for training the network, $20\%$ is reserved for validation, which is used to tune the model and prevent overfitting, and the remaining $20\%$ is set aside for testing, allowing us to evaluate the final performance of the trained model on unseen data.

\begin{table}[t]
\caption{Parameter ranges for Model A, defined according to the \texttt{XSPEC} convention.}
\centering
\begin{tabular}{llcc}
\hline
 & \textbf{Parameter} & \textbf{Range} & \textbf{Distribution} \\
\hline
\multicolumn{4}{c}{\textbf{Model A:} \texttt{tbabs $\times$ powerlaw}} \\
\hline
 & $N_H$          & [0.1, 0.5]    & Uniform \\
 & Photon Index   & [0.5, 3.0]    & Uniform \\
 & NormPL         & [0.1, 10]    & Log Uniform \\
\hline
\end{tabular}
\tablefoot{$N_H$ represents the equivalent hydrogen column density (in units of $10^{22}$ atoms cm$^{-2}$). Photon Index is the photon index of the power law model, and NormPL is its normalization at $1$ keV (photons/keV/cm$^2$/s).}
\label{Tab:model_A}
\end{table}

\subsection{Training of the networks}\label{sec:network}

Predicting continuous-valued physical parameters is a classical statistical regression problem. To address it, we explored three different neural network architectures: an Artificial Neural Network (ANN), a Convolutional Neural Network combined with an ANN (CNN+ANN), and a Gated Recurrent Unit (GRU)-based architecture. Below, we briefly describe these architectures and their relevance for this task:

\begin{itemize}
    \item Artificial Neural Network (ANN): ANNs are the fundamental type of neural network, consisting of multiple layers of neurons interconnected through learnable weights. Each neuron performs a weighted sum of its inputs, followed by a non-linear activation function. ANNs are highly versatile and can approximate complex non-linear relationships, making them suitable for basic regression tasks~\citep{Bishop_1995}.
    \item Convolutional Neural Network (CNN): CNNs are designed to automatically and adaptively learn spatial hierarchies of features from input data through a series of convolutional layers~\citep{Lecun_1998}. The CNN+ANN combination starts with convolutional layers that capture local patterns in the spectral data, which are then flattened and passed to a dense ANN structure to map these features to the final output. This architecture leverages local feature extraction and global pattern recognition, enhancing performance when input data has spatial or temporal dependencies.
    \item Gated Recurrent Unit (GRU): GRU is a type of recurrent neural network (RNN) designed to handle sequential data by maintaining hidden states over time steps~\citep{Cho_2014}. This property allows GRUs to capture temporal dependencies and contextual information in the spectral data, which can be crucial for identifying correlations between spectral bins. GRUs are a simplified version of Long Short-Term Memory (LSTM) networks, providing similar performance but with a lower computational cost, making them a particularly suitable tool in case of limited computational resources.
\end{itemize}

To effectively capture uncertainties in our predictions, we use MCD~\citep{Gal_2015}, a technique that extends traditional dropout to approximate Bayesian inference in neural networks. Dropout is typically employed as a regularization method during training, where a fraction of neurons in each layer is randomly deactivated to prevent overfitting~\citep{Srivastava_2014}. In our approach, however, dropout remains active during inference, enabling the model to generate a range of predictions for each input. This method effectively turns the network into a probabilistic model by simulating posterior sampling. Specifically, when performing inference with dropout enabled, the network generates slightly different predictions for the same input each time, as different subsets of neurons are activated on each forward pass. These varying predictions can then approximate a posterior distribution on the model's output, enabling the neural network to behave like a probabilistic model.

We refer to this approach as MonteXrist (Monte Carlo X-Ray Inference with Spectral Training), leveraging MCD to produce posterior distributions for model parameters. MonteXrist thus provides a robust framework for parameter estimation and uncertainty quantification, with the network generating a distribution of possible outputs that reflects the posterior distribution of model parameters. This probabilistic output enables more reliable predictions by representing the inherent uncertainty in the dataset.

\begin{figure}
    \centering
    \includegraphics[width=1.0\linewidth]{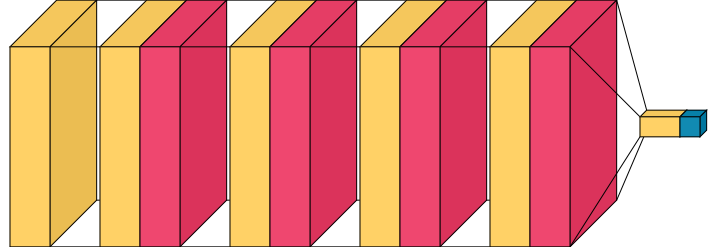}
    \caption{Model architecture used for training, in the case of ANN and GRU models. The yellow blocks represent either GRU layers or ANN layers, depending on the chosen architecture. The red blocks correspond to Dropout layers modified to remain active during inference, enabling posterior distribution sampling for uncertainty quantification. The final blue block is a Dense layer that maps the learned features to the output parameters, such as spectral model components.}
    \label{fig:show_network}
\end{figure}

For the training, we used the \texttt{TensorFlow} and \texttt{Keras} frameworks~\citep{Abadi_2015, Chollet_2015}. In neural networks, certain settings, known as hyperparameters, need to be chosen in advance to define the network's structure and how it learns. These include aspects such as the number of hidden layers, which are layers of neurons that process data and learn patterns within it, and the number of neurons in each layer, which affects the level of detail the network can capture. More layers and more neurons allow the network to model more complex relationships, but they also increase the computational requirements and the risk of overfitting, i.e., when the model becomes too specialized in the training data and performance worsens with new data. Another key hyperparameter is the learning rate, which controls how quickly the model updates its parameters as it learns from data. A higher learning rate makes the model adjust faster but can lead to instability, while a lower rate makes learning slower but often more precise. We also set the batch size, which is the number of samples the network processes at a time before updating its parameters; this choice influences both the speed and stability of learning. The architecture of the network, including the arrangement of GRU (or ANN) layers, Dropout layers, and the layer used for output regression, is illustrated in Fig.~\ref{fig:show_network}.

Due to resource constraints, we did not perform an exhaustive search for the optimal combination of hyperparameters. Instead, we chose a standard set of values for each type of network, as detailed in Table~\ref{Tab:hyperparameters_A}. We tested if varying the number of neurons in each layer (rather than using the same number in all layers) could improve performance, but we found no substantial benefit from this approach.

The model was trained over a maximum of $5000$ cycles, called epochs. An epoch refers to one complete pass through the entire training dataset, during which the network adjusts its parameters based on the data. We used a technique called early stopping to prevent overfitting. Early stopping monitors the model’s performance on a separate validation dataset, and if there is no improvement over a certain number of epochs (known as patience), training stops early to save computational resources. In our case, we set a patience of $100$ epochs and began monitoring after the $50^{\mathrm{th}}$ epoch, which gives the network time to start learning meaningful patterns~\citep{Goodfellow_2016}.

The loss function used for training was the Mean Squared Error (MSE), which is a common choice for regression problems. In general, a loss function is a mathematical measure of how well the model’s predictions match the true values in the training data. The loss function guides the model's learning process by providing feedback on the quality of predictions. The MSE, specifically, is defined as:
\begin{equation} 
\mathrm{MSE} = \frac{1}{N} \sum_{i=1}^{N} (y_i - \hat{y}_i)^2, 
\end{equation} 
where $y_i$ represents the true value, $\hat{y}_i$ the predicted value, and $N$ is the number of samples. The MSE was minimized using the \textit{Adam} optimizer~\citep{Kingma_2014}, a variant of stochastic gradient descent that adapts the learning rate for each parameter based on estimates of the first and second moments of the gradients. \textit{Adam} is well-suited for problems with noisy gradients and sparse data, making it a robust choice for our setup.

\begin{table}[t]
\caption{Hyperparameters used in the NNs for Model A.}
\centering
\begin{tabular}{llccc}
\hline
 & \textbf{Parameter} & \textbf{ANN} & \textbf{CNN+ANN} & \textbf{GRU} \\
\hline
\multicolumn{5}{c}{\textbf{Model A:} \texttt{tbabs $\times$ powerlaw}} \\
\hline
 & Hidden Layers                & 4    & 2+4      & 4    \\
 & Neurons per Layer            & 100  & 100+100  & 100  \\
 & Dropout                      & 10\% & 10\%     & 10\% \\
 & Learning Rate $(10^{-4})$    & 1    & 1        & 1    \\
 & Batch Size                   & 32   & 32       & 32   \\
\hline
\end{tabular}
\label{Tab:hyperparameters_A}
\end{table}

\subsection{Evaluation of performances}\label{sec:performances}

To evaluate the performance of our neural networks, in addition to the loss function, we considered the following metrics: Mean Squared Logarithmic Error (MSLE), Mean Absolute Error (MAE), and the Coefficient of Determination (R$^2$). These metrics, taken together, allow us to assess different aspects of the model’s performance:

\begin{itemize}
    \item MSLE measures the squared logarithmic differences between the true and predicted values. It focuses on capturing relative errors and reducing the impact of outliers.
    \item MAE computes the mean of the absolute differences between the predicted and true values, offering an intuitive measure of the average error magnitude.
    \item R$\mathbf{^2}$, defined as:
    \begin{equation}
    R^2 = 1 - \frac{\sum_{i=1}^{N}(y_i - \hat{y}_i)^2}{\sum_{i=1}^{N}(y_i - \bar{y})^2},
    \end{equation}
    where $y_i$ are the true values, $\hat{y}_i$ the predicted values, and $\bar{y}$ the mean of $y_i$, quantifies the proportion of variance explained by the model, ranging from 0 to 1. Higher values indicate better fits.
\end{itemize}

To determine which model architecture performs best, we prioritize the following:
\begin{itemize}
    \item A higher R$^2$ value, which indicates the model captures most of the variance in the data.
    \item Lower MSE and MAE values, signaling smaller overall and absolute errors in predictions.
    \item Consistently low MSLE, demonstrating robustness to parameter scaling and avoiding excessive errors in parameters with smaller ranges.
\end{itemize}

For both of the two datasets (the Dataset-short and the Dataset-long), the GRU architecture demonstrated the best performance on all metrics compared to the ANN and CNN+ANN models (see Table~\ref{Tab:metrics}). In particular, its R$^2$ values ($0.995$) and low MSE ($4 \times 10^{-4}$) highlight its ability to achieve both precision and generalization. The differences in MSLE and MAE further confirm the GRU's ability to minimize prediction errors across all parameter scales.

The superior performance of the GRU model can be attributed to its ability to capture sequential dependencies within the spectral data, which exhibit inherent correlations between energy channels. While the ANN and CNN+ANN architectures effectively learn global features from the spectra, they do not explicitly model the sequential relationships between neighboring bins. The GRU, on the other hand, is a type of recurrent neural network designed to maintain and update hidden states through time steps, allowing it to capture these dependencies effectively.

In particular, GRUs excel at retaining important information over longer sequences while mitigating the vanishing gradient problem that can occur in traditional RNNs. This ability to propagate information through sequential steps allows the GRU to better model the continuous nature of the X-ray spectra, where each energy bin may be influenced by its neighbors. As a result, the GRU can more accurately predict the source emission's physical parameters, which are encoded across the entire spectrum.

Moreover, GRUs are computationally more efficient than other recurrent architectures like LSTMs, as they have fewer gates and require fewer resources to train while maintaining a strong capability to capture long-term dependencies. This makes them particularly well-suited to tasks like X-ray spectral fitting, where clear patterns and correlations across the spectrum need to be understood holistically.

\begin{table}[t]
\caption{Evaluation of performances for Model A for the two different datasets.}
\centering
\begin{tabular}{llccc}
\hline
 & \textbf{Metric} & \textbf{ANN} & \textbf{CNN+ANN} & \textbf{GRU}\\
\hline
\multicolumn{5}{c}{\textbf{Dataset-short}} \\
\hline
 & MSE ($10^{-3}$) & 1.8   & 2.0      & 0.4   \\
 & MSLE ($10^{-4}$)& 8.2   & 9.1      & 1.7   \\
 & MAE ($10^{-2}$) & 3.0   & 3.2      & 1.2   \\
 & R$^2$           & 0.981 & 0.978    & 0.995 \\
\hline
\multicolumn{5}{c}{\textbf{Dataset-long}} \\
\hline
 & MSE ($10^{-3}$) & 2.0   & 1.9      & 0.4   \\
 & MSLE ($10^{-4}$)& 8.8   & 8.7      & 1.6   \\
 & MAE ($10^{-2}$) & 3.2   & 3.1      & 1.2   \\
 & R$^2$           & 0.979 & 0.980    & 0.995 \\
\hline
\end{tabular}
\tablefoot{The performance is evaluated using the test dataset.}
\label{Tab:metrics}
\end{table}

\begin{figure}[t]
    \centering
    \includegraphics[width=1\linewidth]{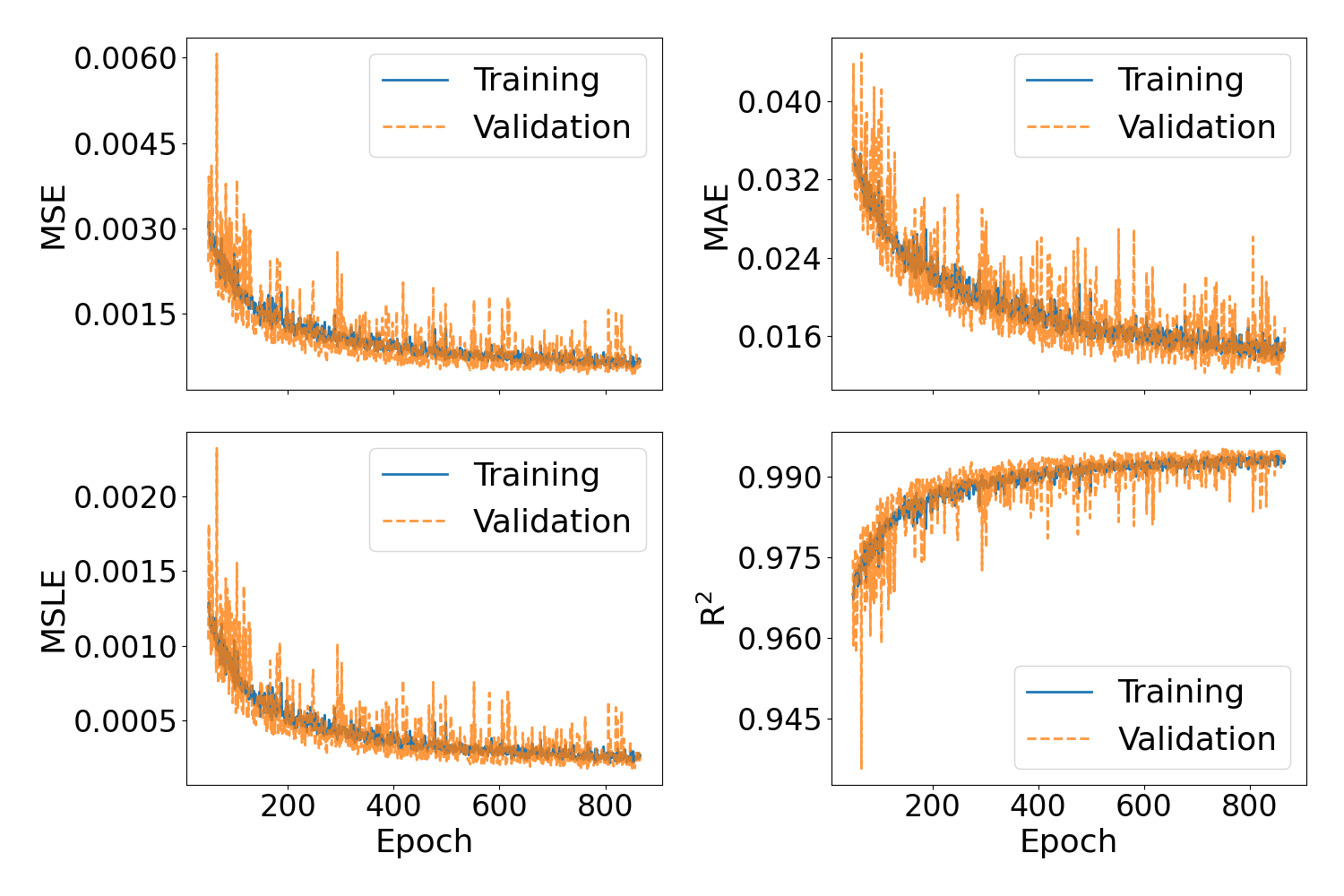}
    \caption{Training history of the neural network using the GRU architecture in the case of Model A on the Dataset-short. Each plot compares the metric on the training set (solid blue line) with its counterpart on the validation set (solid orange line). The training curve represents the model's performance on the data it learns from, while the validation curve indicates its performance on a separate dataset (the validation dataset) during the training phase.}
    \label{fig:loss_curves}
\end{figure}

The training history of the GRU model for Model A on the Dataset-short is depicted in Figure~\ref{fig:loss_curves}, showcasing the evolution of four metrics over $\sim 900$ epochs: the loss function (MSE), MAE, MSLE, and R$^2$. The loss curves demonstrate a clear downward trend for both training and validation datasets, indicating effective minimization of MSE and good generalization without significant overfitting. Similar trends are observed for MAE and MSLE, with errors reducing as epochs progress. R$^2$ scores steadily increase, approaching values close to 1, indicating the model's competence in capturing target parameter variance. The minimal gap between training and validation metrics confirms robust convergence and high accuracy, indicating successful model training.

\begin{figure*}[h]
    \centering
    \begin{subfigure}{0.33\textwidth}
        \includegraphics[width=\linewidth]{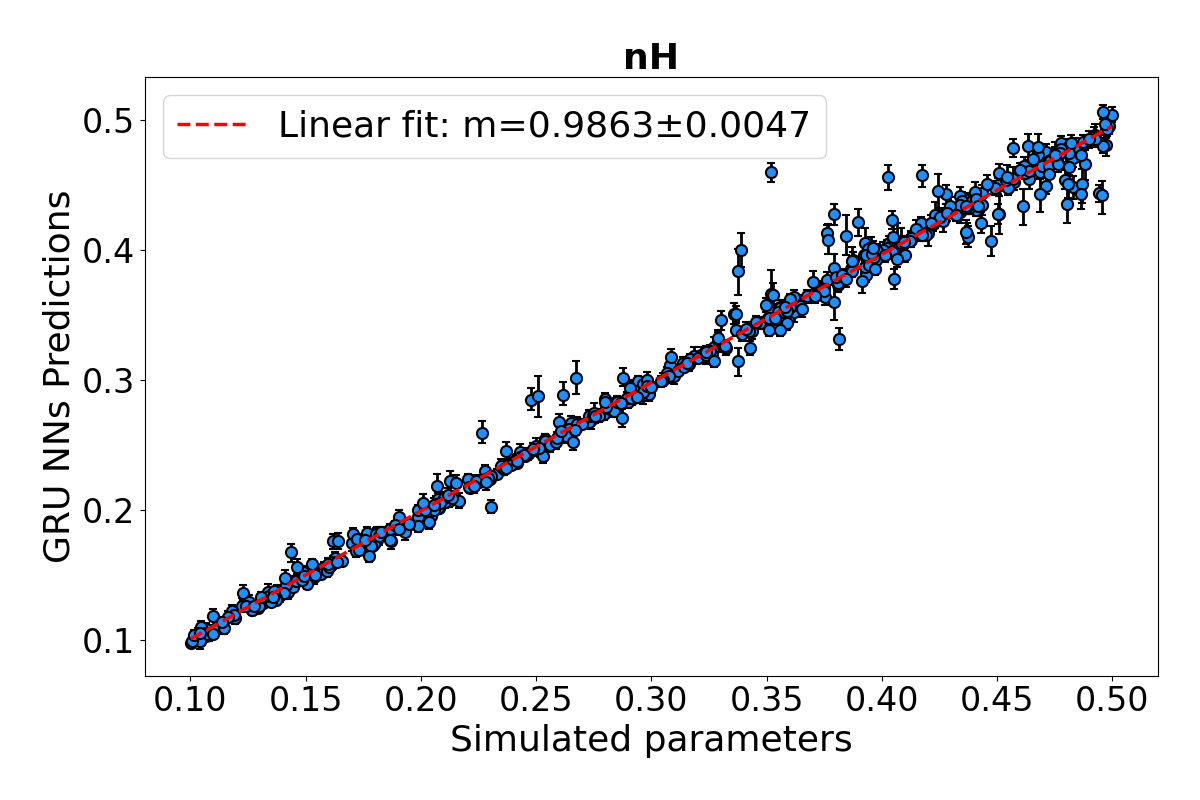}
    \end{subfigure}
    \begin{subfigure}{0.33\textwidth}
        \includegraphics[width=\linewidth]{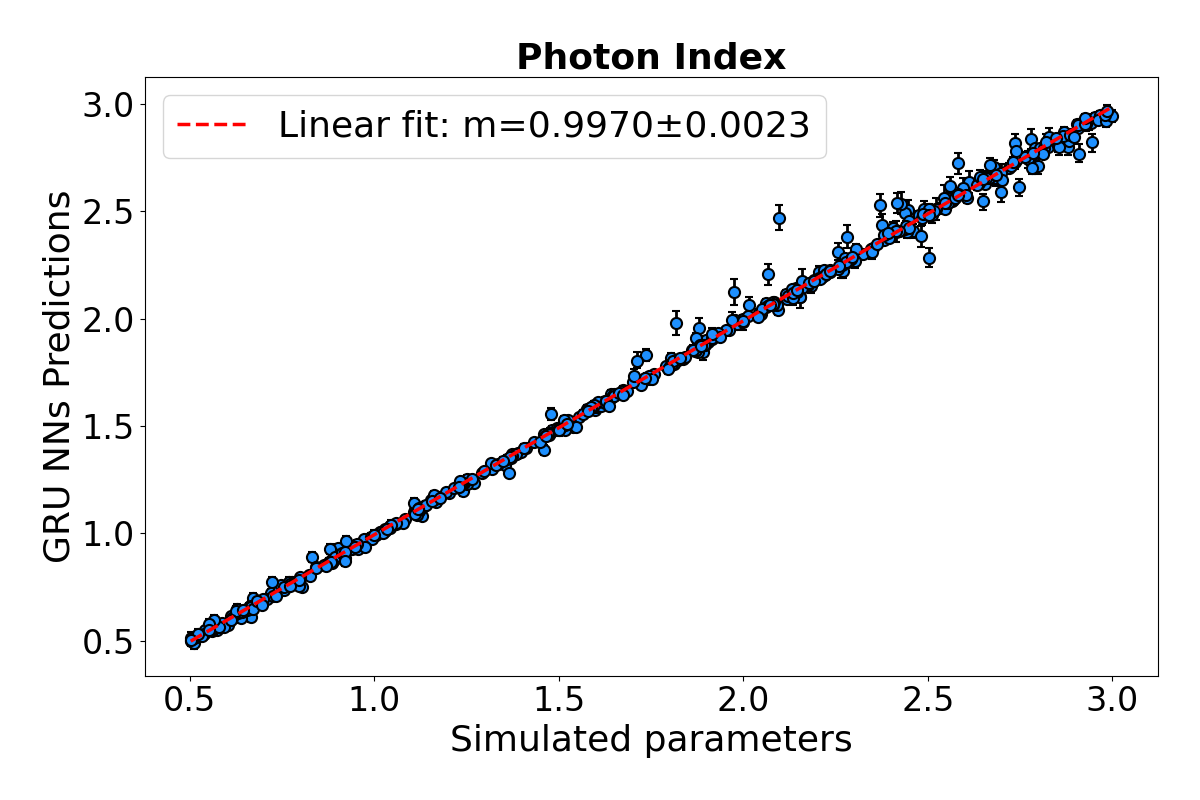}
    \end{subfigure}
    \begin{subfigure}{0.33\textwidth}
        \includegraphics[width=\linewidth]{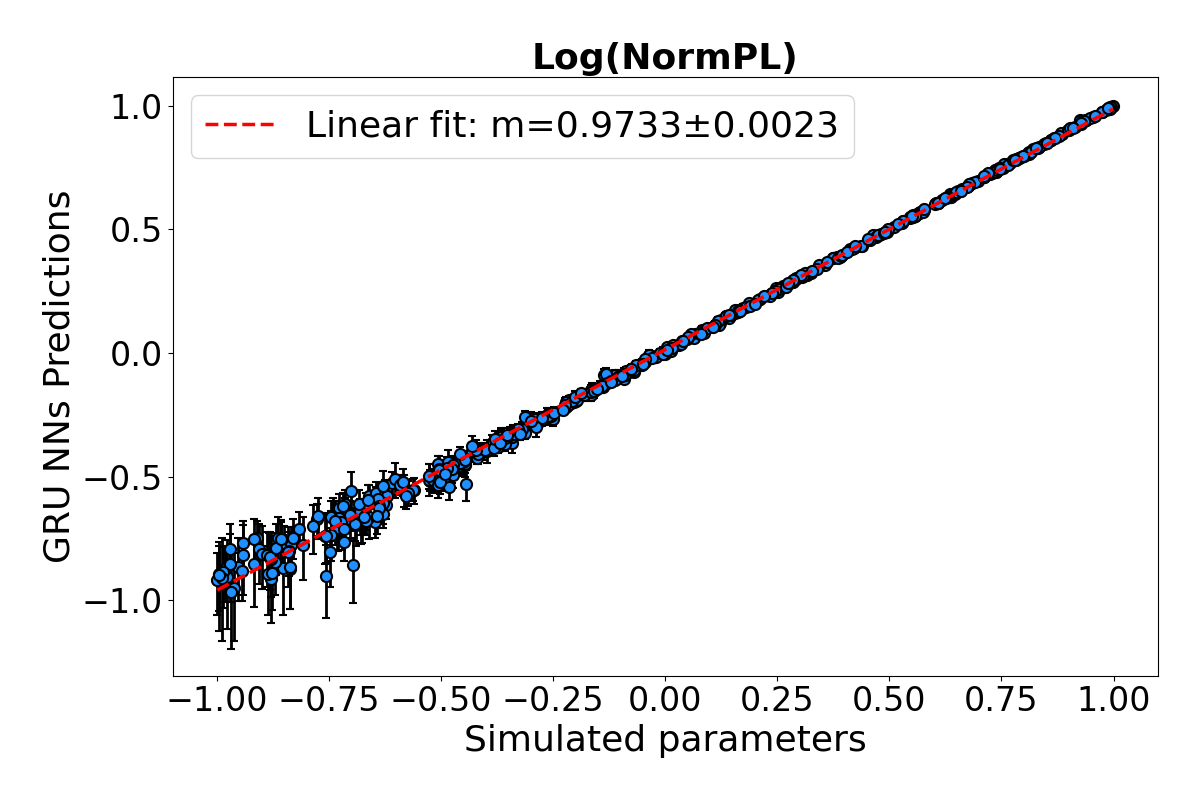}
    \end{subfigure}
    
    \begin{subfigure}{0.33\textwidth}
        \includegraphics[width=\linewidth]{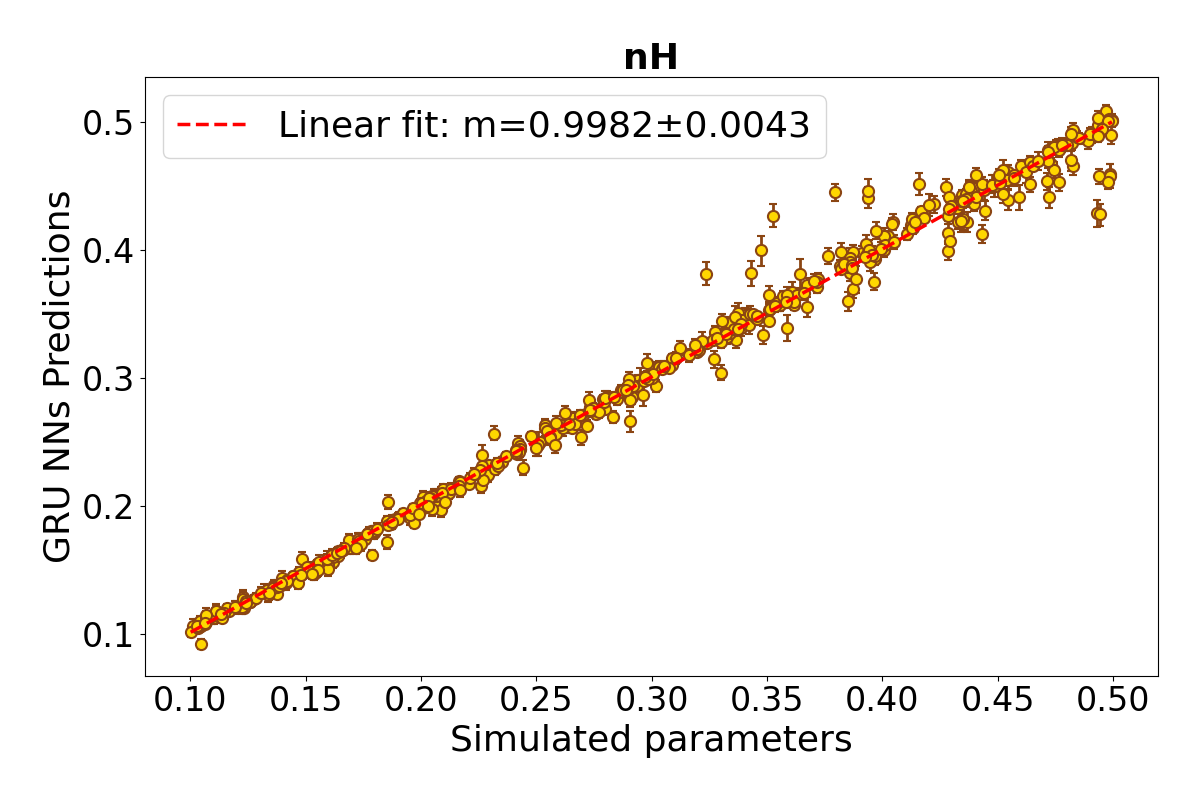}
    \end{subfigure}
    \begin{subfigure}{0.33\textwidth}
        \includegraphics[width=\linewidth]{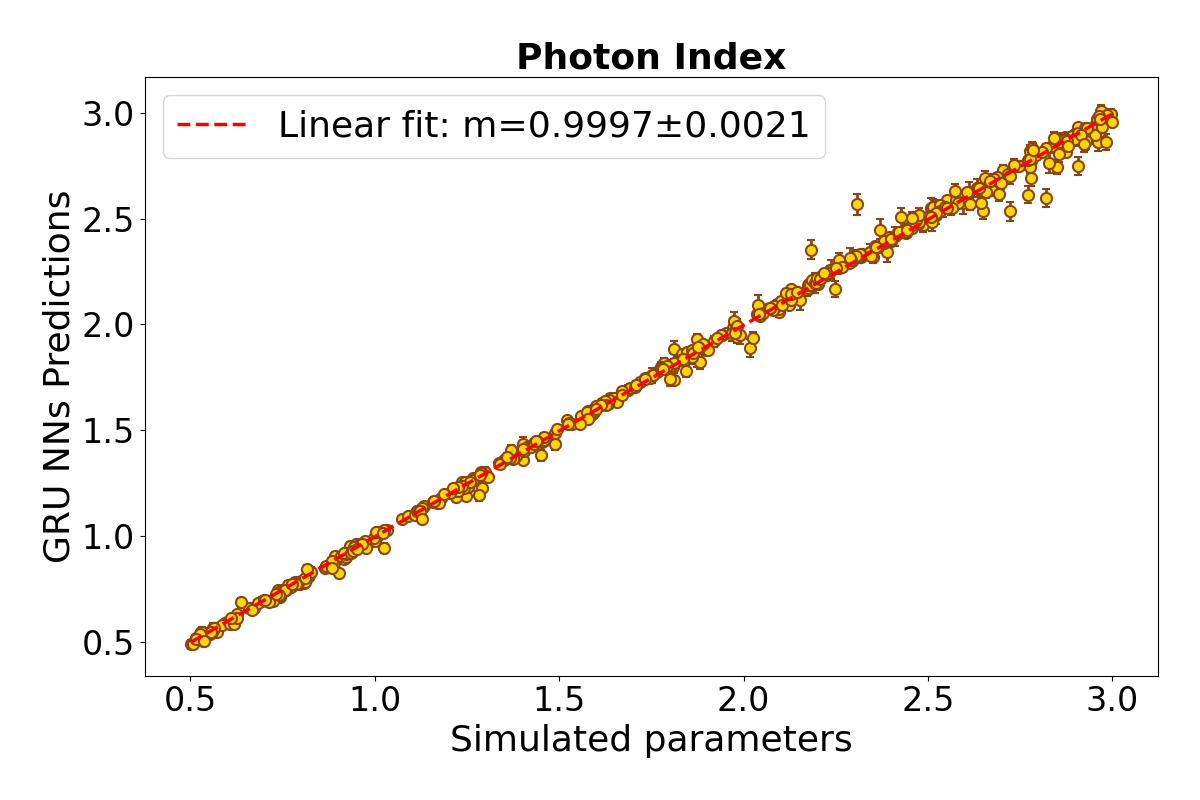}
    \end{subfigure}
    \begin{subfigure}{0.33\textwidth}
        \includegraphics[width=\linewidth]{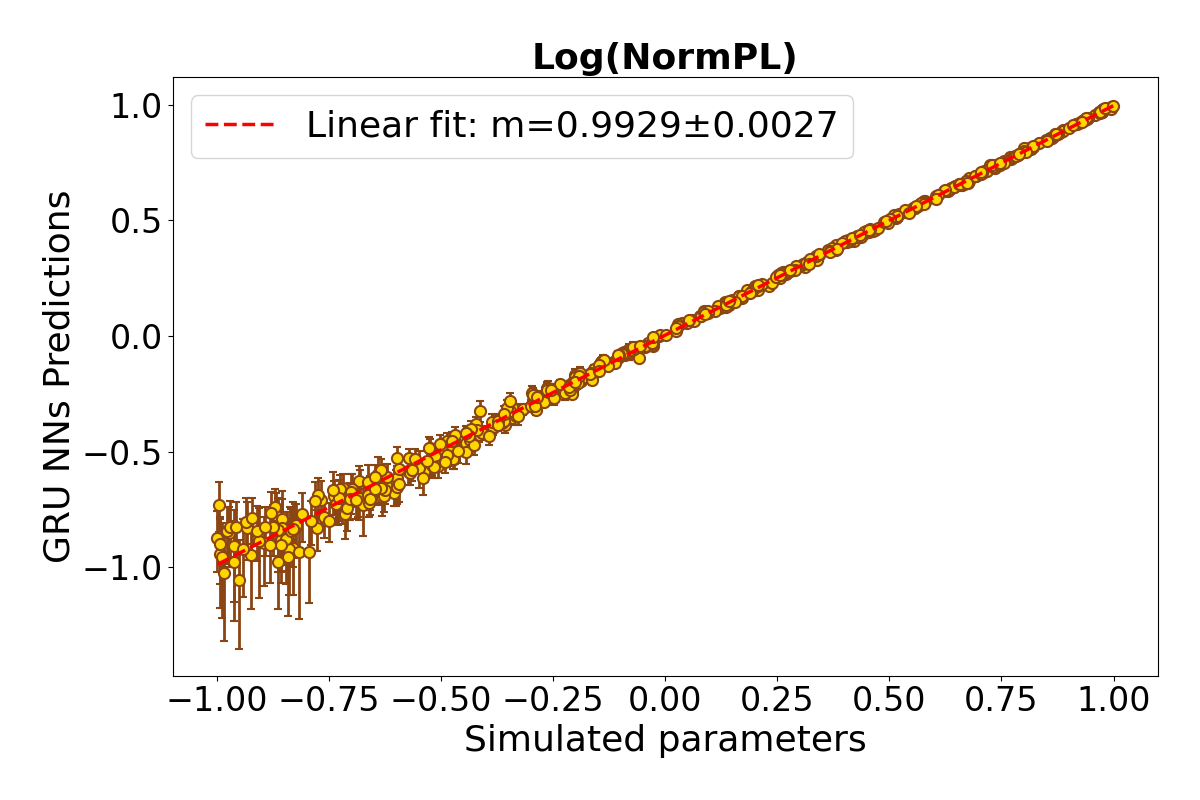}
    \end{subfigure}
    
    \caption{Inferred model parameters are compared with input model parameters for two cases: Dataset-short (top row) and Dataset-long (bottom row). Posteriors were calculated for $500$ test spectra from the test dataset. The medians of the posteriors were derived from $1000$ samples for each of the $500$ test spectra, and the error on the median was determined from the 68\% quantile of the distribution. A linear regression coefficient was calculated for each parameter across the $500$ test samples.}
    \label{fig:parameters}
\end{figure*}

To visually inspect the results from the network, we generated scatter plots using $500$ randomly selected simulated spectra from the test dataset, where the x-axis shows the true parameter values and the y-axis the values predicted by the GRU network (see Fig.~\ref{fig:parameters}). A linear regression fit of the points reveals a good agreement between input and output parameters, with the linear regression coefficient being very close to $1$ for both the Dataset-short (upper row) and Dataset-long (lower row).

\begin{figure*}[h]
    \centering
    \begin{subfigure}{0.49\textwidth}
        \includegraphics[width=\linewidth]{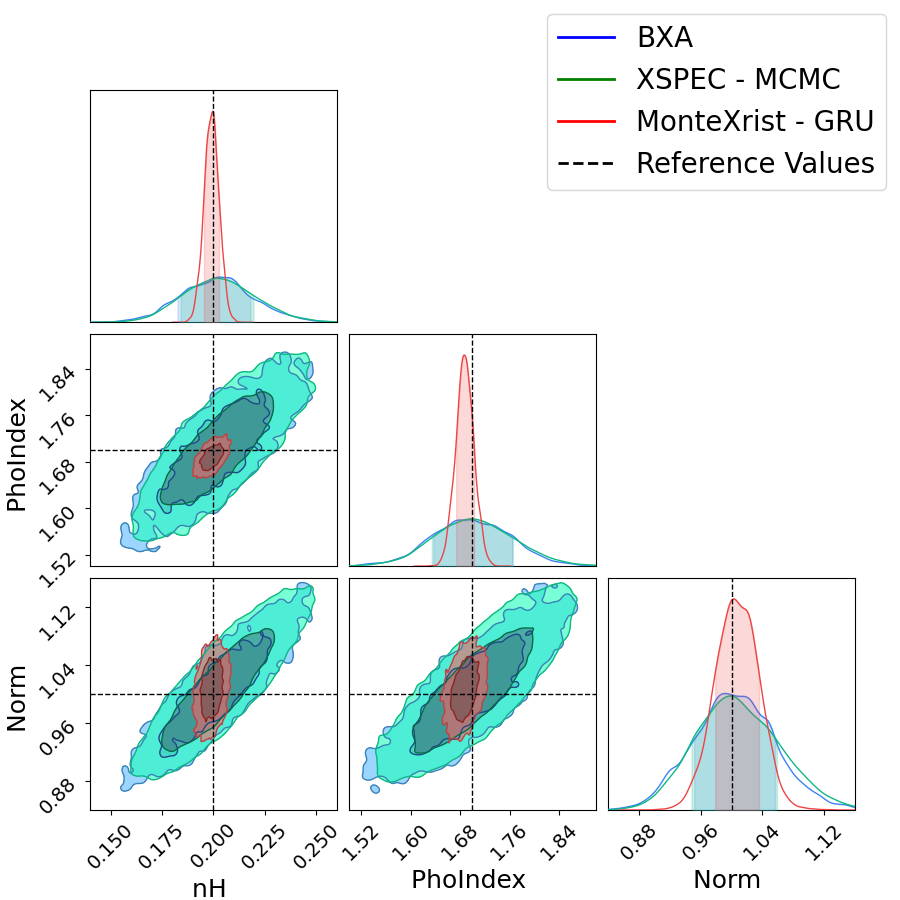}
    \end{subfigure}
    \begin{subfigure}{0.49\textwidth}
        \includegraphics[width=\linewidth]{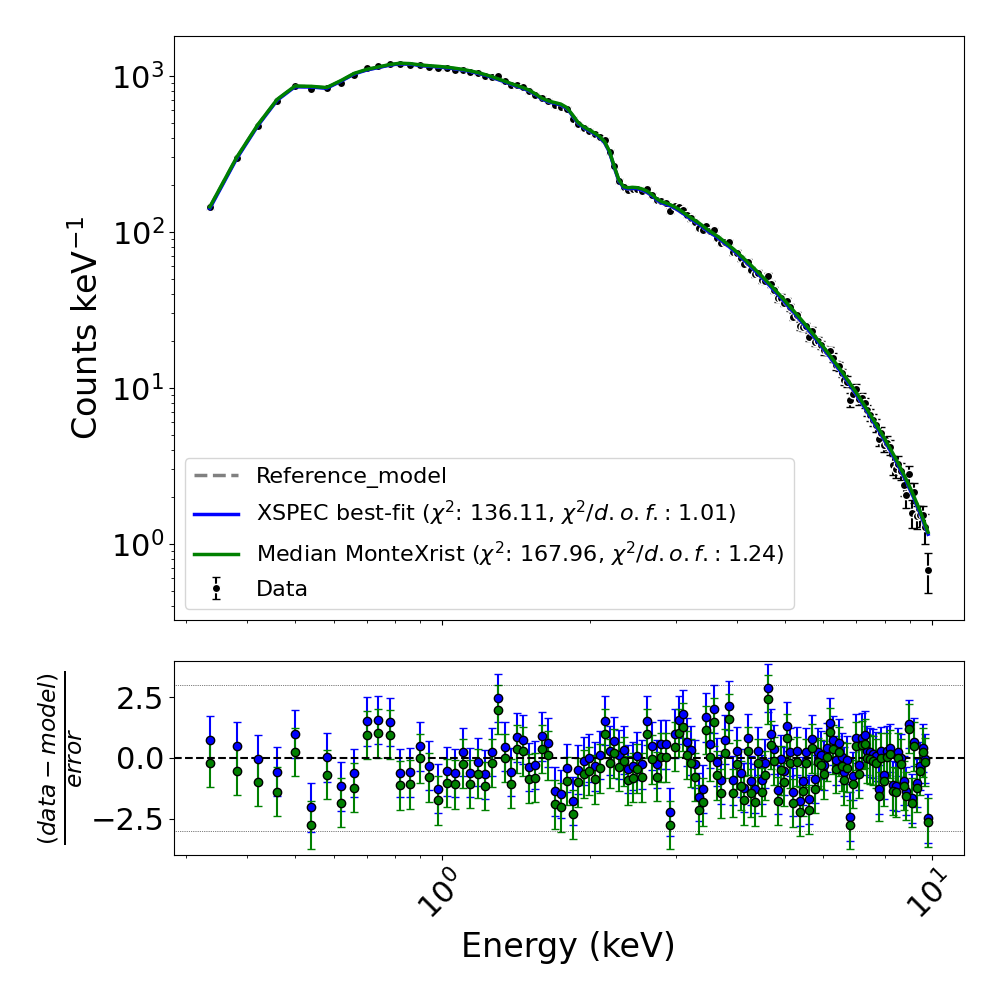}
    \end{subfigure}
    \caption{Left: The posterior distribution estimated for the reference spectrum with $2000$ counts (red). The posterior distributions inferred from Bayesian fits with BXA and XSPEC are also shown in blue and green, respectively. Right: The emission spectrum and residuals corresponding to the reference Model A with $2000$ counts, together with the folded best-fit model from GRU (green solid line) and XSPEC (blue dashed line) with a frequentist approach.}
    \label{fig:corner_pwl_1}
\end{figure*}

\begin{figure*}[h]
    \centering
    \begin{subfigure}{0.49\textwidth}
        \includegraphics[width=\linewidth]{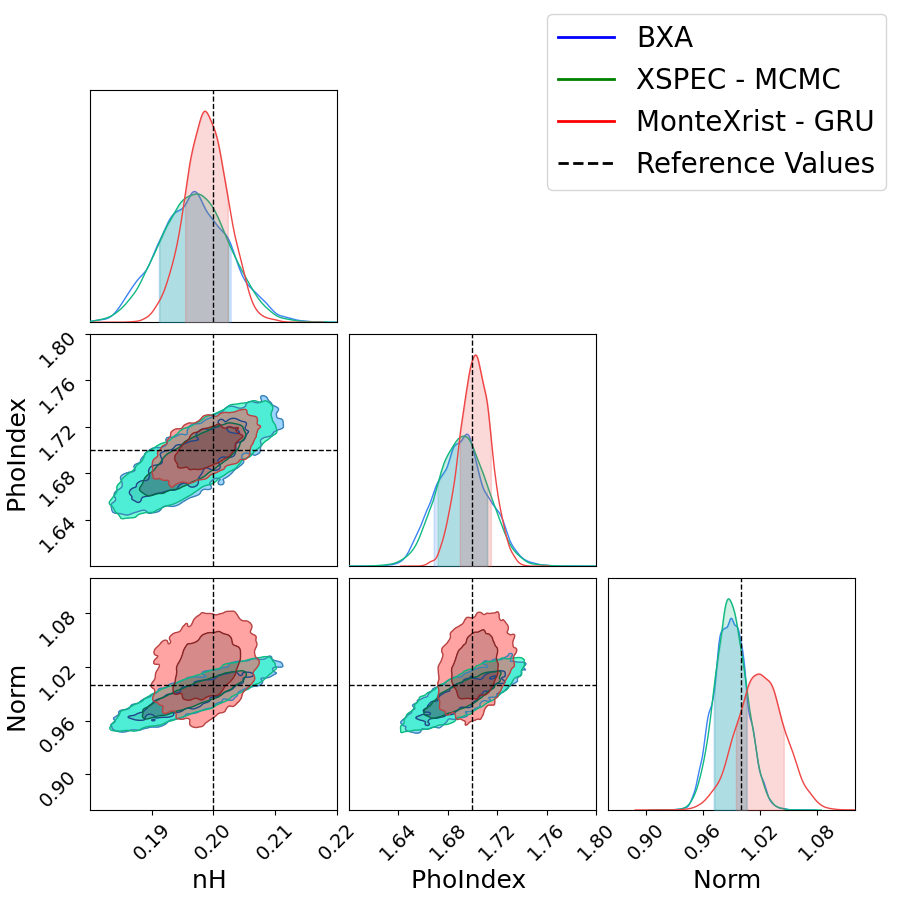}
    \end{subfigure}
    \begin{subfigure}{0.49\textwidth}
        \includegraphics[width=\linewidth]{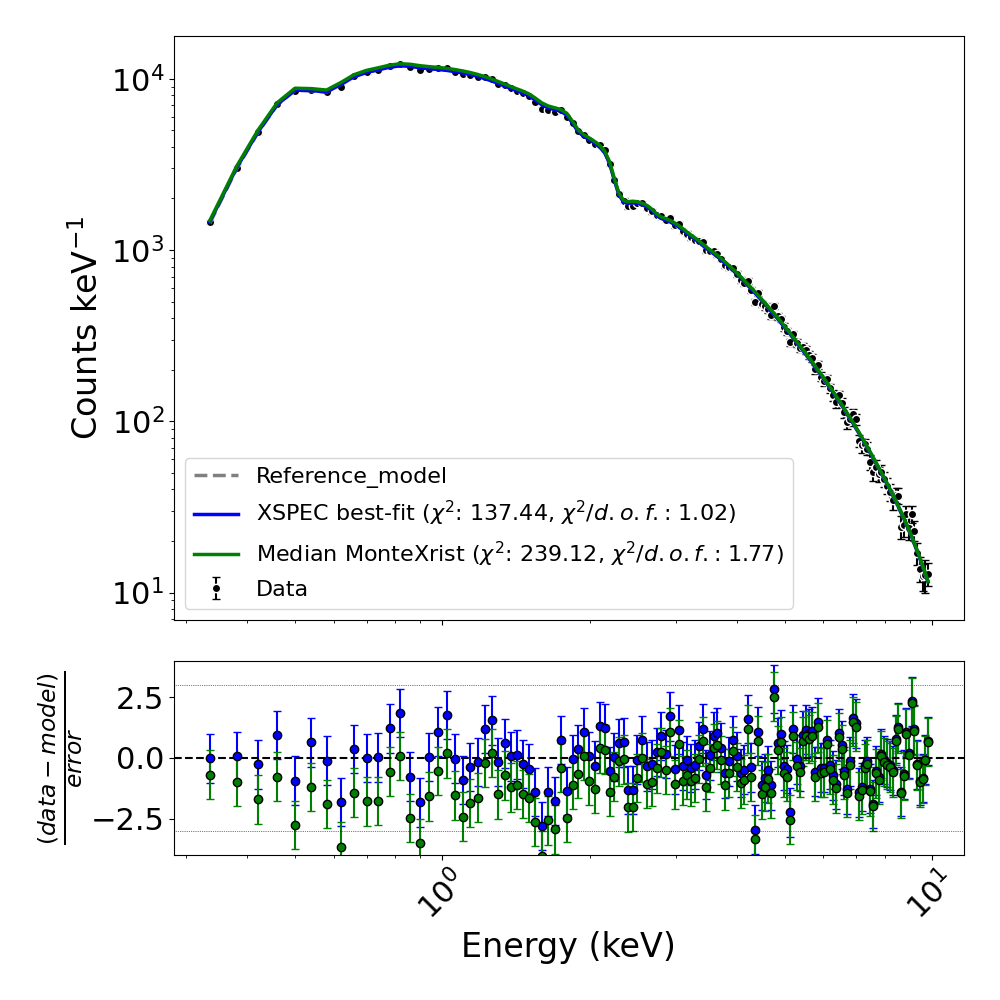}
    \end{subfigure}
    \caption{Same as Figure~\ref{fig:corner_pwl_1} but for the reference spectrum with $2 \times 10^4$ counts.}
    \label{fig:corner_pwl_10}
\end{figure*}

We derive the posterior distributions for the reference spectra for the cases of $2 \times 10^3$ and $2 \times 10^4$ counts, comparing these with the posterior distributions obtained from \texttt{XSPEC} with Bayesian inference enabled and from \texttt{BXA}, a tool that combines \texttt{XSPEC} with nested sampling techniques for Bayesian analysis. For both methods, we adopted the same priors listed in Tab.~\ref{Tab:model_A} for Model A, and \texttt{BXA} was run using its default parameter settings. These comparisons are illustrated in Figure~\ref{fig:corner_pwl_1} and~\ref{fig:corner_pwl_10}. We find a good overall agreement in both the best-fit parameters and the posterior distributions. Our method yields a notably narrower posterior distribution in the scenario with fewer counts ($2 \times 10^3$) than the other two techniques. However, some degeneracy is seen in the normalization parameter (Norm). In the higher-count scenario ($2 \times 10^4$ counts), the distributions are similar. This degeneracy in the power law normalization might result from adjustments in the photon index, producing a less precise posterior for this parameter. This indicates that our method performs well even with limited statistics, consistent with the results in Table~\ref{Tab:metrics}. Although we did not conduct a systematic study as it is beyond the scope of the paper, we observed that the dropout rate influences both the posterior distributions and network performance. As expected, higher dropout values ($>10\%$), compared with the values given in the Tab.~\ref{Tab:hyperparameters_A}, lead to broader posterior distributions, increasing uncertainty estimation but also adding noise to parameter recovery. Conversely, lower dropout rates (<5\%) produce narrower posteriors, but at the cost of overfitting, reducing generalization to unseen spectra.
Moreover, the effect of dropout is strongly influenced by other hyperparameters, such as the number of neurons per layer and the learning rate, which can impact how the network compensates for the regularization induced by dropout.
These outcomes confirm the importance of choosing an appropriate dropout rate to balance uncertainty estimation and model stability. A more systematic exploration could further refine this choice.

In Figures~\ref{fig:corner_pwl_1} and~\ref{fig:corner_pwl_10}, we compare the best-fit model obtained from \texttt{XSPEC} using the frequentist approach (blue line) and the median prediction from the GRU-based MonteXrist approach (green line) with the reference spectral data (black points). The fit residuals for both methods are shown in the lower panels, along with their respective $\chi^2 / \mathrm{d.o.f.}$ values.

For the $2 \times 10^3$ counts reference spectrum, Fig.~\ref{fig:corner_pwl_1}, both methods provide a reasonable fit to the data, with the \texttt{XSPEC} best-fit achieving a $\chi^2 / \mathrm{d.o.f.}$ (degrees of freedom) value of $1.01$, while the MonteXrist approach shows a slightly higher $\chi^2 / \mathrm{d.o.f.}$ value of $1.24$. Despite the marginally worse statistical fit, the GRU-based MonteXrist approach performs well in capturing the overall spectral shape, and the residuals between the two methods are comparable.
Similar conclusions can be drawn for the $2 \times 10^4$ counts reference spectrum (Fig.~\ref{fig:corner_pwl_10}). Here, both methods are again found to be comparable, ensuring the method's robustness.

\section{Real data scenario\label{sec:real_data}}

After demonstrating the effectiveness of our technique on simulated data, we verified its applicability to real data. This step is crucial in machine learning applications, as it determines the model's ability to generalize and operate effectively on data not seen during training. 
We chose the same NICER observation of 4U 1820-30 analysed by~\citet{Barret_2024} to facilitate model comparison. The analysis procedure for this source closely follows that of~\citet{Barret_2024} and is detailed in the following subsection. In the context of this study, we will focus exclusively on analyzing the spectrum of the persistent X-ray emission.

\subsection{Data reduction}

Among low mass x-ray binaries (LMXBs), 4U 1820-30 stands out due to its unique characteristics. 
It has the shortest known orbital period of any LMXB at just $11.4$ minutes \citep{Stella_1987}, classifying it as an Ultra-Compact X-ray Binary (for a review, see \citealt{Armas_2023}). Additionally, it belongs to a subclass known as X-ray bursters, characterized by periodic, rapid increases in luminosity, called Type I X-ray bursts, caused by thermonuclear burning of hydrogen and heavier elements at the NS surface. 

4U 1820-30 has been observed multiple times by NICER. To compare our results with those obtained by \cite{Barret_2024}, we analyzed the same observation referenced by the authors, specifically the one from October 28, 2017 (ObsID \texttt{1050300108}), with a total exposure time of $29$ ks. We processed the NICER data using the {\tt nicerl2} pipeline tool in NICERDAS v12, available with HEASoft v6.34, following the recommended calibration procedures and standard screening (the calibration database version used was {\tt xti20240206}).
We generated the light curve in the $0.3-10$ keV range using {\tt nicerl3-lc} to check for Type-I X-ray bursts, identifying one. To focus on analyzing the persistent spectrum, we excluded the burst by selecting a time interval outside of it. Specifically, we extracted the spectrum from a time window of $190$ seconds, starting $200$ seconds before and ending $10$ seconds prior to the burst, using {\tt nicerl3-spect} and excluding detectors $14$ and $34$ due to their elevated noise levels. We used the {\tt scorpeon} background file with the {\tt bkgformat=file} option for background subtraction. Following the NICER team recommendations, we applied a systematic error of $1.5 \%$ and grouped the data with optimal rebinning ({\tt grouptype=optmin}) with the additional requirement of a minimum number of $25$ counts per energy bin \citep{Kaastra_2016}. 

\subsection{Training and predictions}
The NICER spectrum can be described by a combination of a power law component and a black-body emission, {\tt bbodyrad} in XSPEC \citep{Arnaud_1996}, both corrected for interstellar medium absorption using the model {\tt tbabs}. The \texttt{bbodyrad} normalization is directly related to the radius of the emission region  ${\rm R_{bb}}$ through the formula: $\rm{K_{bb}=(R_{bb}/D_{10\, kpc})}^{2}$, where ${\rm D_{10\, kpc}}$ is the source distance in units of 10 kpc. The chosen model, referred to as Model B, is therefore: \texttt{tbabs $\times$ (powerlaw + bbodyrad)}.

\begin{figure}
    \centering
    \includegraphics[width=1\linewidth]{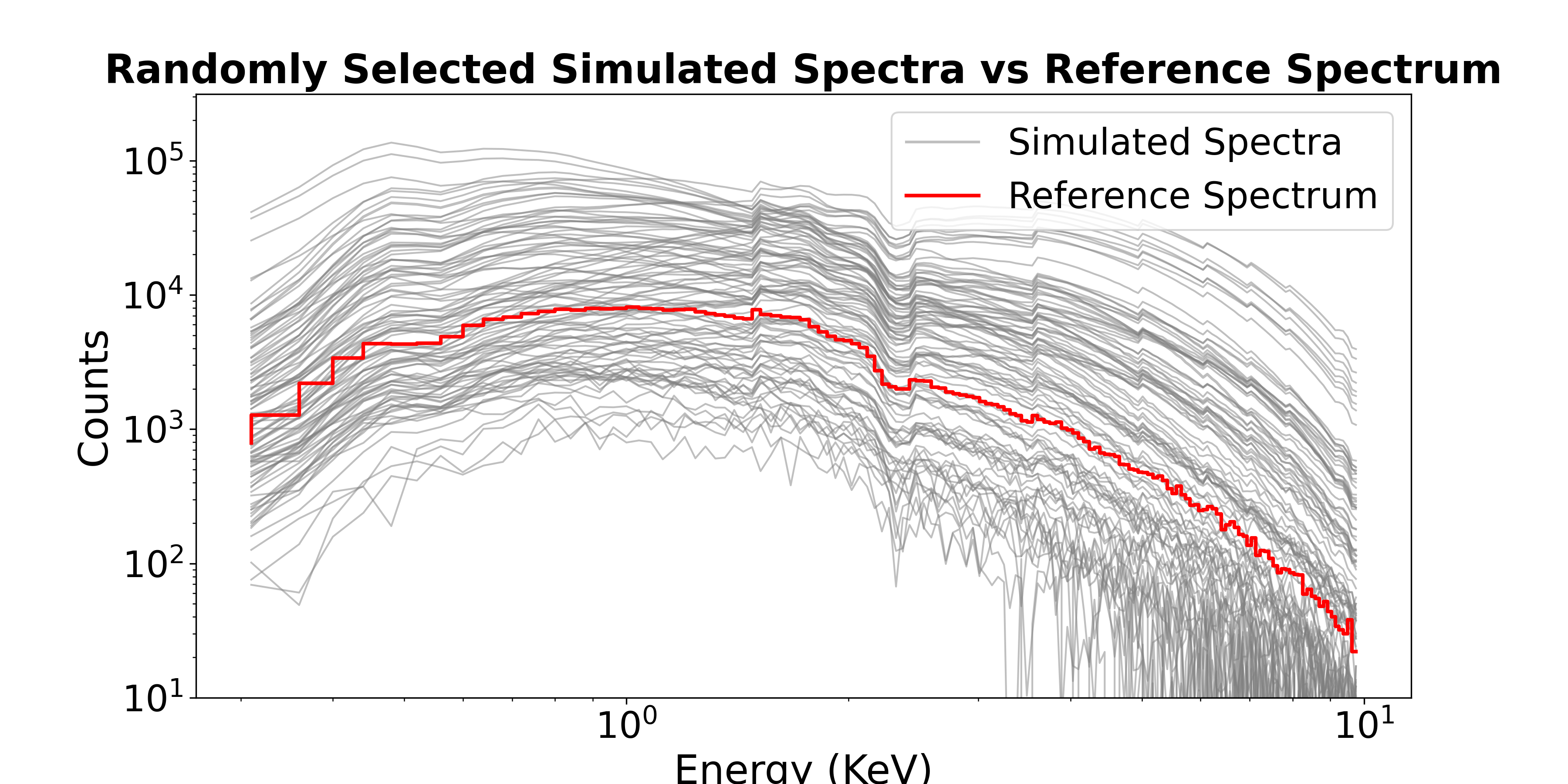}
    \caption{Emission spectrum from 4U 1820-30 (red line) and $100$ randomly selected simulated (Model B) spectra from the prior distribution (grey lines).}
    \label{fig:dataset_bbodyrad}
\end{figure}

The first step is, therefore, to train a neural network with this specific model to ensure that it learns the relevant parameter relationships. 
We generated a dataset of $3 \times 10^5$ simulated spectra to accomplish this, following the same procedure outlined in the previous section. The ranges of the model parameters explored in these simulations are reported in Table~\ref{Tab:model_B}. For reference, we show in Figure~\ref{fig:dataset_bbodyrad} the source emission spectrum with $100$ randomly selected simulated spectra. As in our earlier tests, both the input spectra and the corresponding output parameters were preprocessed. Specifically, we standardized the simulated spectra and normalized the model parameters using the methods described in Section~\ref{sec:dataset_generation}. This preprocessing step was essential to ensure stable training and improve the convergence of the neural network models.

\begin{table}[t]
\caption{Parameter ranges for Model B.}
\centering
\begin{tabular}{llcc}
\hline
 & \textbf{Parameter} & \textbf{Range} & \textbf{Distribution} \\
\hline
\multicolumn{4}{c}{\textbf{Model B:} \texttt{tbabs $\times$ (powerlaw + bbodyrad)}} \\
\hline
 & $N_H$          & [0.15, 0.35]    & Uniform \\
 & Photon Index   & [1.0, 3.0]    & Uniform \\
 & NormPL         & [0.1, 10]    & Log Uniform \\
 & kT             & [0.3, 3.0]    & Uniform \\
 & NormBB         & [10, 1000]    & Log Uniform \\
\hline
\end{tabular}
\tablefoot{$N_H$ represents the equivalent hydrogen column density (in units of $10^{22}$ atoms cm$^{-2}$). Photon Index is the photon index of the power-law model, NormPL is its normalization at $1$ keV (photons/keV/cm$^2$/s), kT denotes the blackbody temperature in keV, and NormBB is its normalization, expressed as $R^2_{km}/D^2_{10}$, where $R_{km}$ is the source radius in km and $D_{10}$ is the distance to the source in units of $10$ kpc.}
\label{Tab:model_B}
\end{table}

\begin{table}[t]
\caption{Hyperparameters used in the NNs for Model B.}
\centering
\begin{tabular}{llccc}
\hline
 & \textbf{Parameter} & \textbf{ANN} & \textbf{CNN+ANN} & \textbf{GRU} \\
\hline
\multicolumn{5}{c}{\textbf{Model B:} \texttt{tbabs $\times$ (powerlaw + bbodyrad)}} \\
\hline
 & Hidden Layers                & 4    & 2+4      & 4    \\
 & Neurons per Layer            & 200  & 200+200  & 200  \\
 & Dropout                      & 5\%  & 5\%      & 5\%  \\
 & Learning Rate $(10^{-4})$    & 0.1  & 0.1      & 0.1  \\
 & Batch Size                   & 32   & 32       & 32   \\
\hline
\end{tabular}
\label{Tab:hyperparameters_B}
\end{table}

\begin{table}[ht]
\caption{Evaluation of results for Model B.}
\centering
\begin{tabular}{llccc}
\hline
 & \textbf{Metric} & \textbf{ANN} & \textbf{CNN+ANN} & \textbf{GRU}\\
\hline
 & MSE ($10^{-3}$) & 2.8   & 3.3      & 2.2   \\
 & MSLE ($10^{-3}$)& 1.4   & 1.7      & 1.1   \\
 & MAE ($10^{-2}$) & 3.0   & 3.3      & 2.0   \\
 & R$^2$           & 0.968 & 0.963    & 0.975 \\
\hline
\end{tabular}
\label{Tab:metrics_bbodyrad}
\end{table}

We again tested three different neural network architectures: ANN, CNN+ANN, and GRU, using the hyperparameters listed in Table~\ref{Tab:hyperparameters_B}. Also in this case, see Table~\ref{Tab:metrics_bbodyrad}, the GRU architecture consistently outperforms the others, making it the best choice for this task. We proceeded with the GRU model.

\begin{figure}
    \centering
    \includegraphics[width=1.0\linewidth]{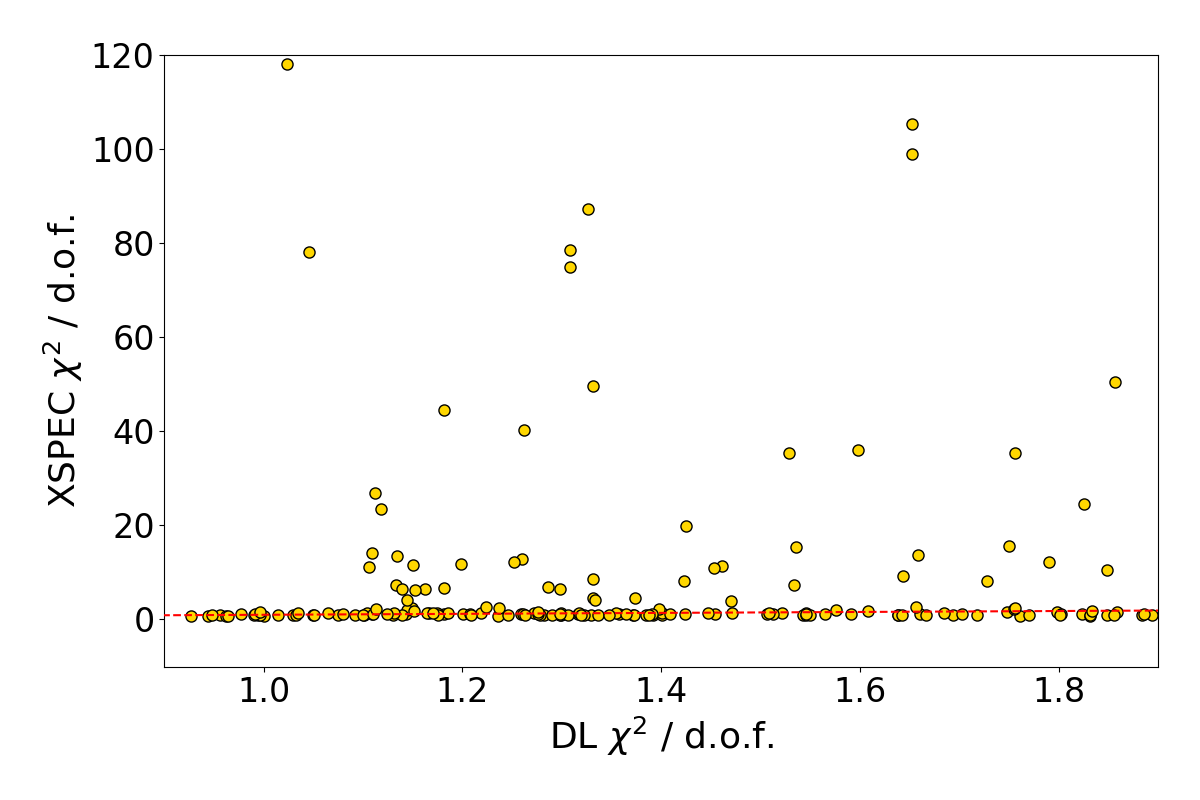}
    \caption{Comparison of XSPEC spectral fitting ($\chi^2$ minimization) and predictions from the trained GRU neural network for the Model B. The red dashed line represents the expected 1:1 correlation. Points deviating significantly from this line indicate cases where XSPEC became trapped in a local minimum, highlighting the robustness of the neural network approach in avoiding false minima.}
    \label{fig:single_stat}
\end{figure}

\begin{figure*}[h]
    \centering
    \begin{subfigure}{0.33\textwidth}
        \includegraphics[width=\linewidth]{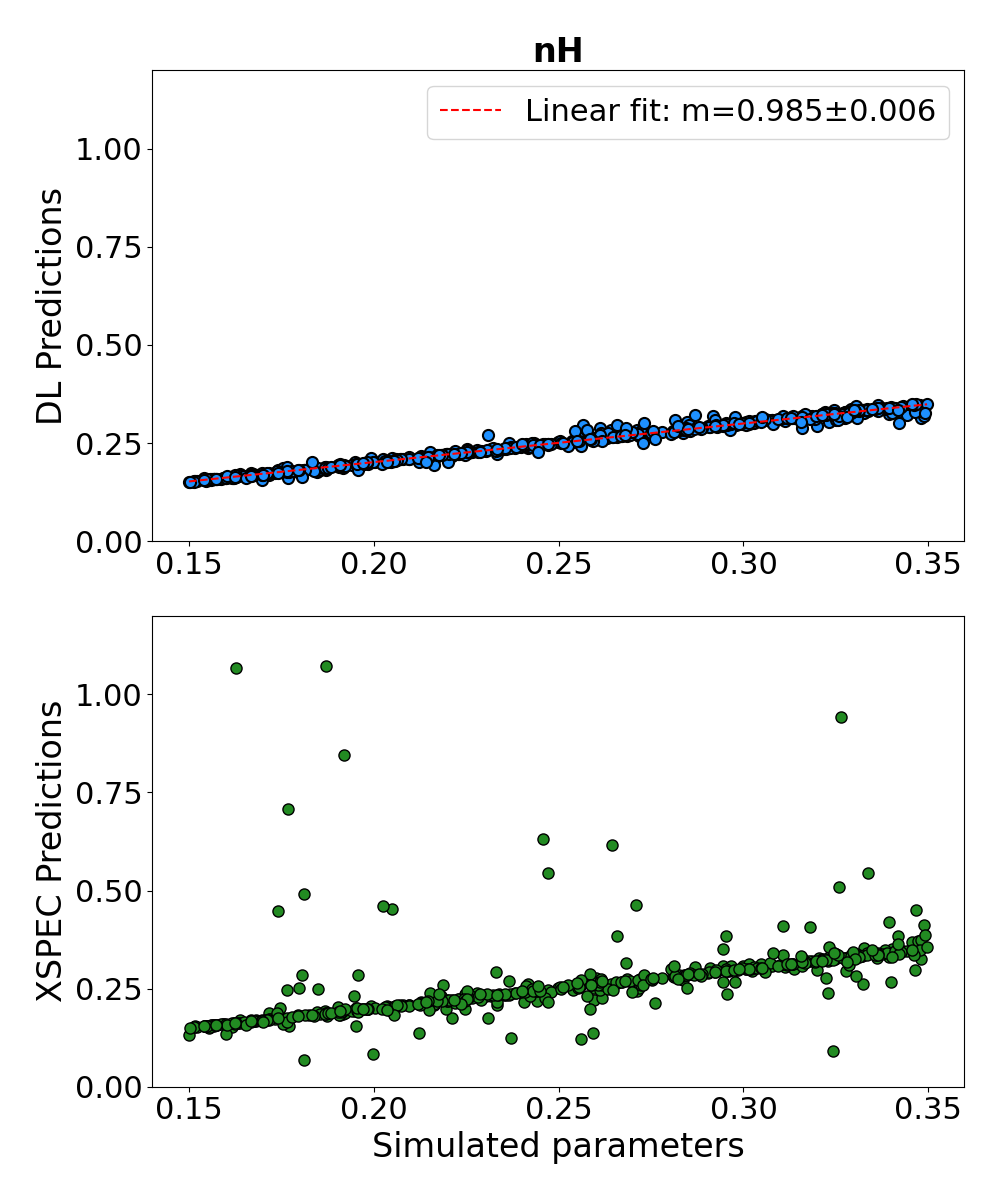}
    \end{subfigure}
    \begin{subfigure}{0.33\textwidth}
        \includegraphics[width=\linewidth]{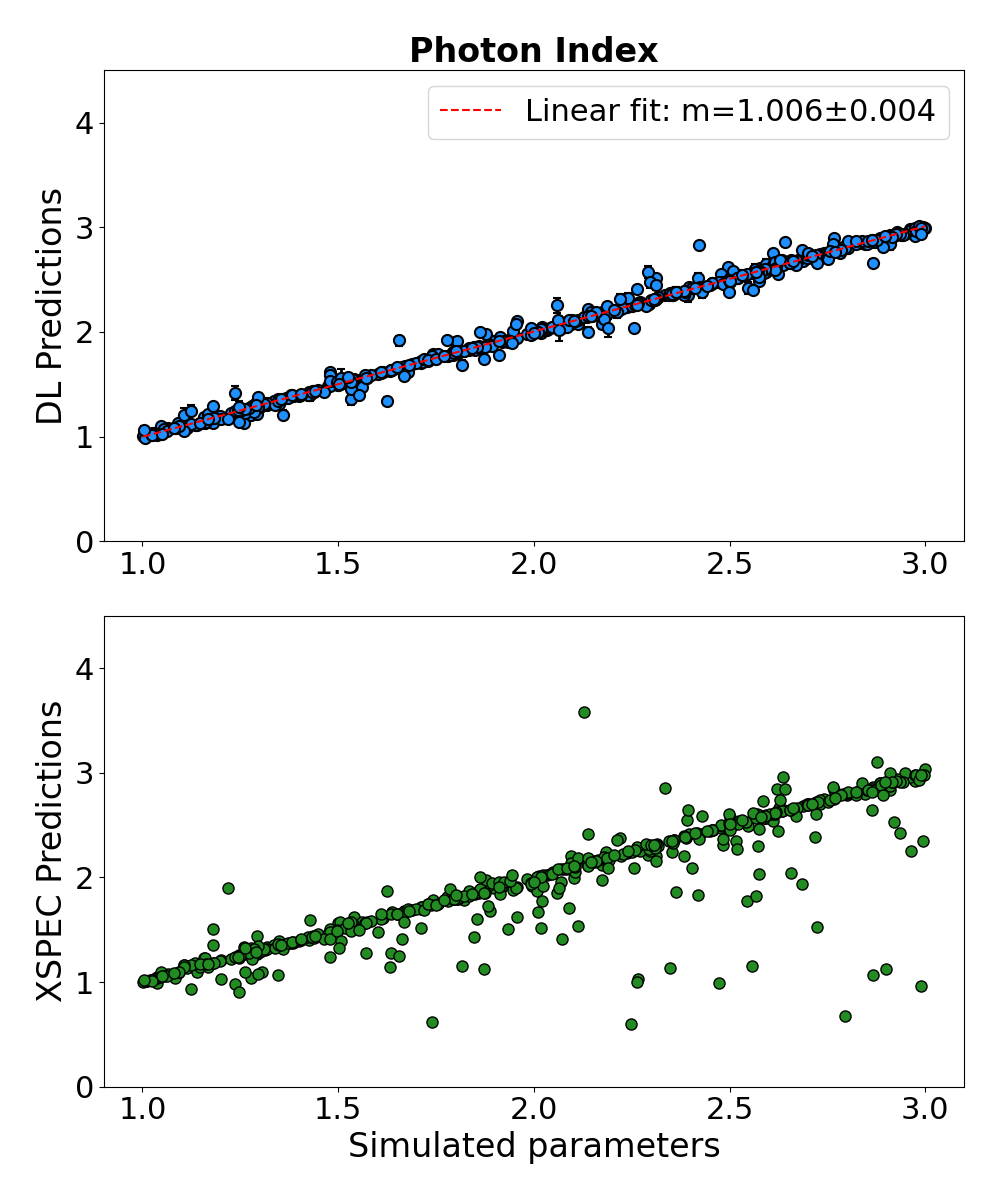}
    \end{subfigure}
    \begin{subfigure}{0.33\textwidth}
        \includegraphics[width=\linewidth]{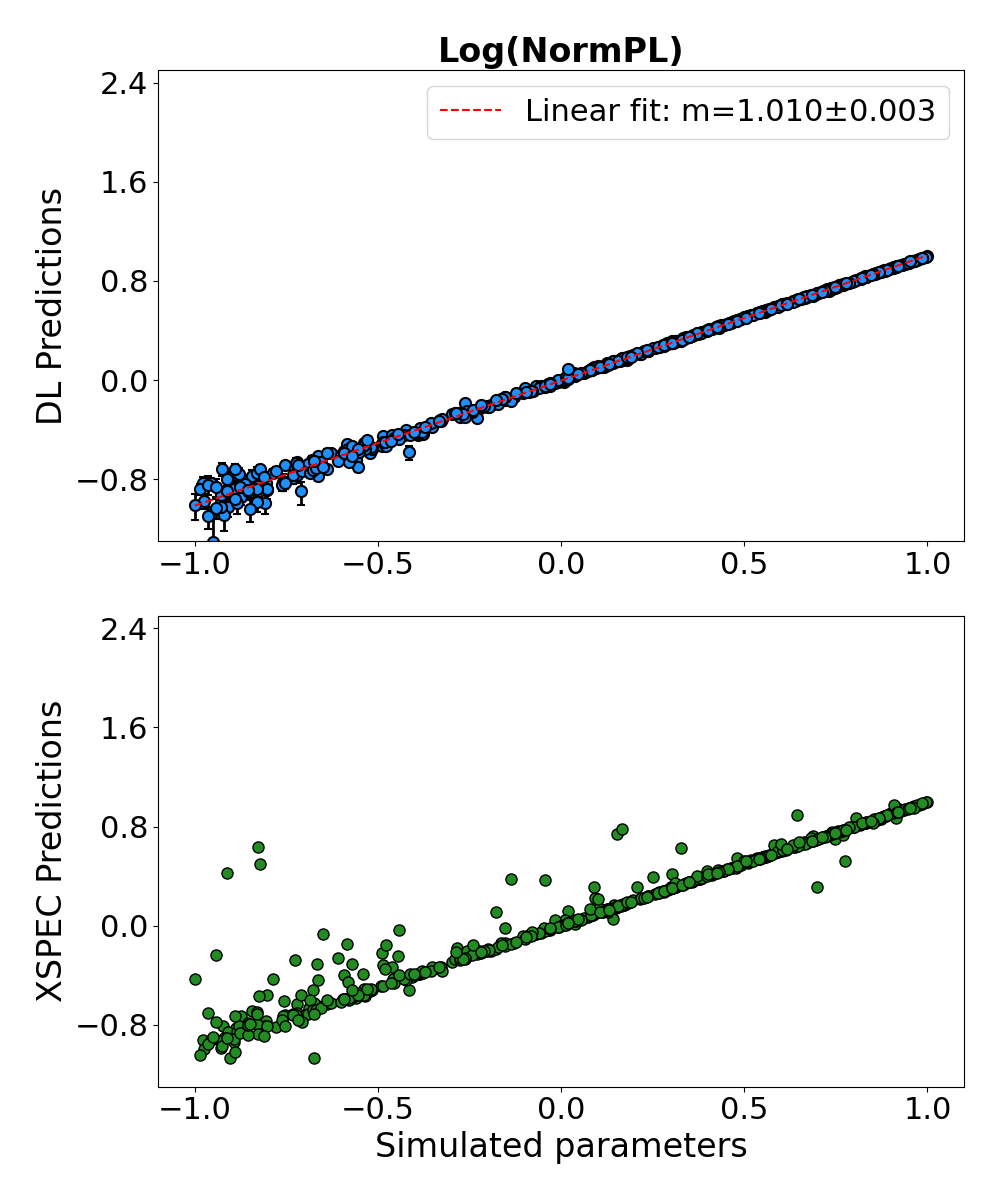}
    \end{subfigure}
    \begin{subfigure}{0.40\textwidth}
        \includegraphics[width=\linewidth]{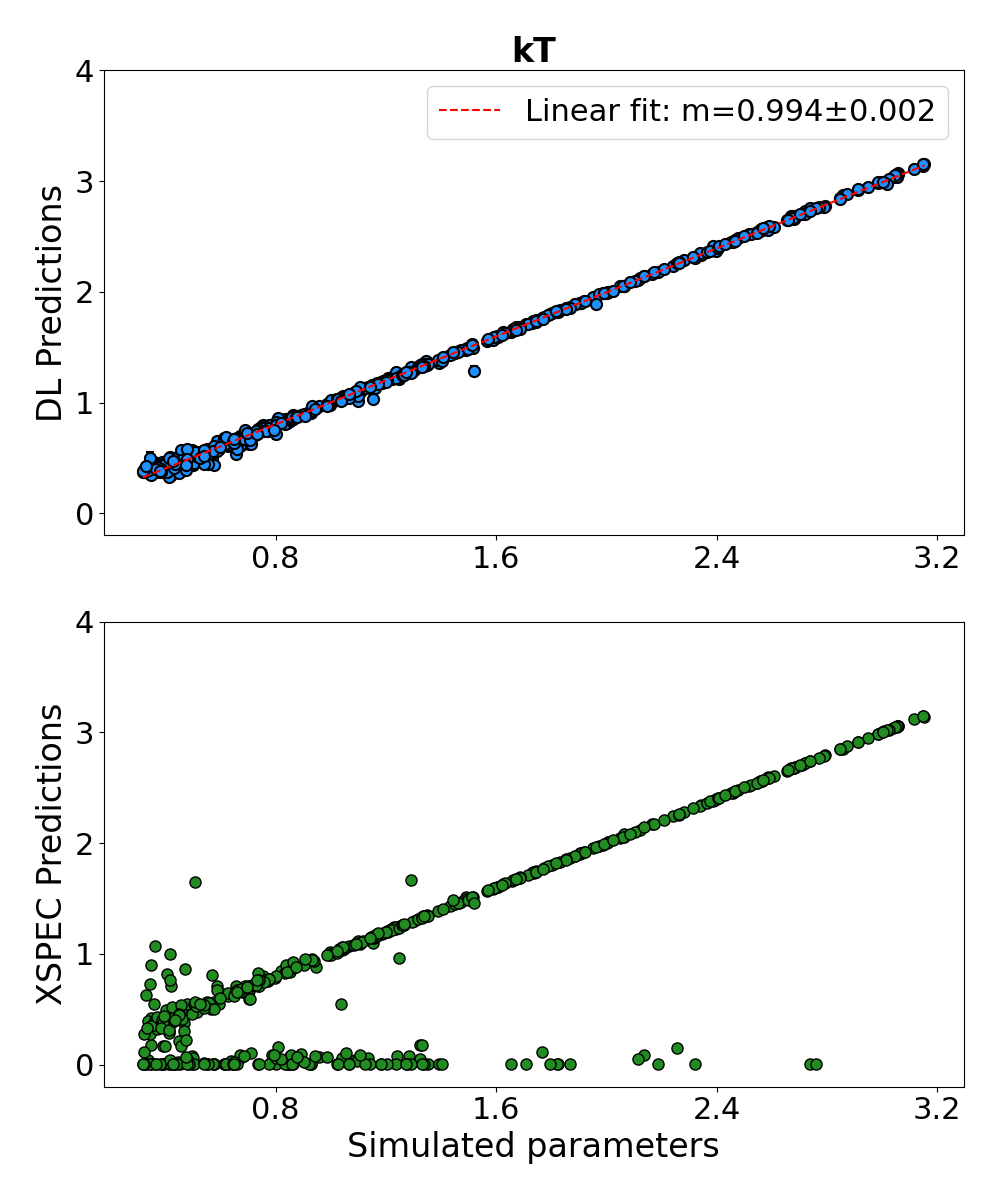}
    \end{subfigure}
    \begin{subfigure}{0.40\textwidth}
        \includegraphics[width=\linewidth]{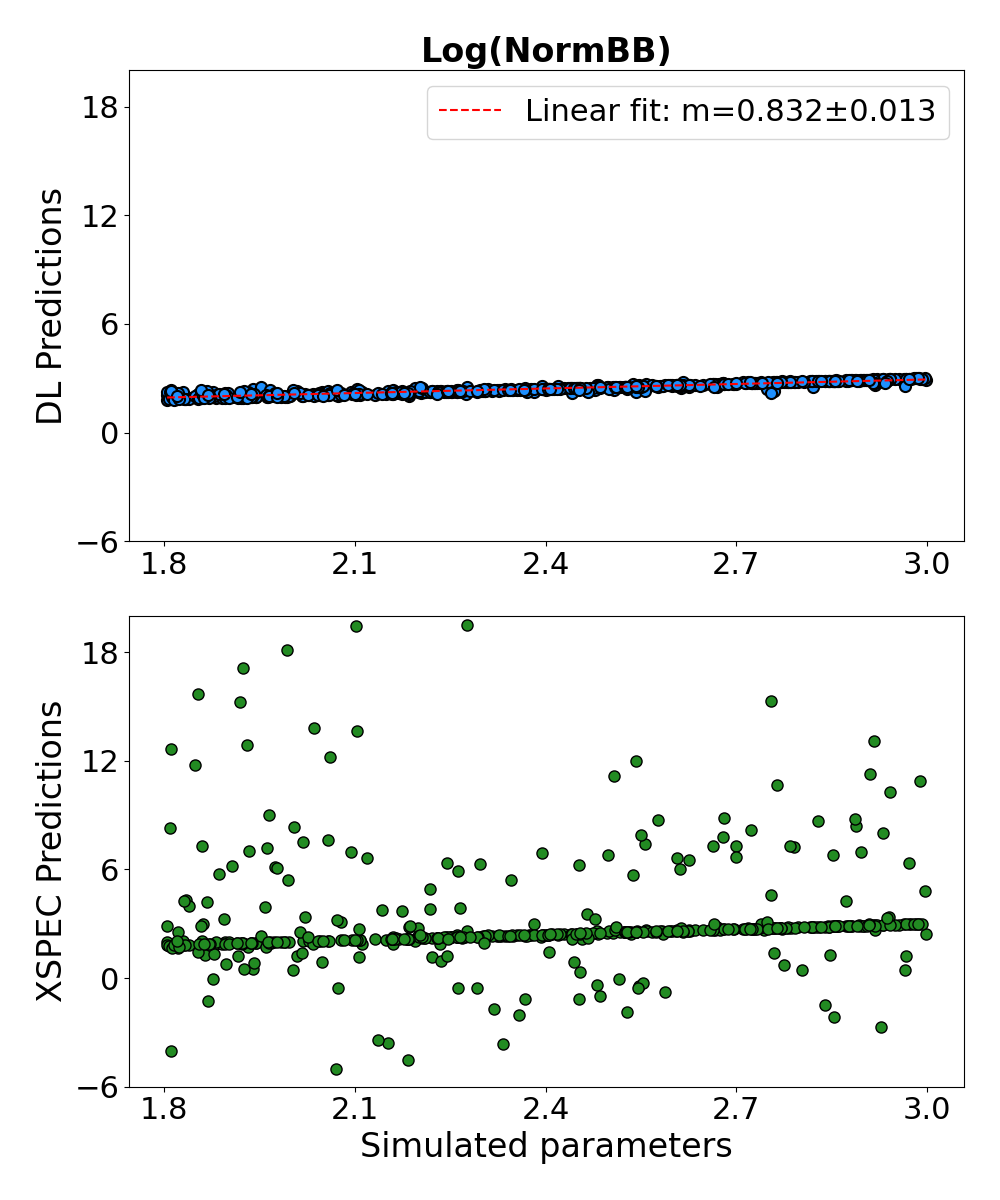}
    \end{subfigure}
    \caption{Similar to Figure~\ref{fig:parameters}, the top panels (blue points) employs the trained GRU neural network, while the bottom panels (green points) uses \texttt{XSPEC} with Bayesian inference disabled, for the Model B.}
    \label{fig:local_minima}
\end{figure*}

In Figure~\ref{fig:single_stat}, we highlight how the \texttt{XSPEC} fitting procedure, when using $\chi^2$ minimization, can become trapped in local minima, leading to significantly incorrect parameter estimates, as previously noted by \citet{Parker_2022}. It is particularly clear from the outliers which deviate from the expected correlation line, shedding light to cases where the optimization fails to find the global minimum.
This effect is further explored in Fig.~\ref{fig:local_minima}, where we compare the parameter estimated with our neural network and those derived from \texttt{XSPEC}. The presence of outliers in the \texttt{XSPEC} results confirms that local minima can lead to inaccurate parameter estimates. In contrast, our GRU-based neural network significantly reduces this risk. While our method does not always achieve the same level of precision as \texttt{XSPEC} when the latter converges correctly, it consistently avoids false minima, ensuring stable and reliable parameter estimates across different cases.


\begin{figure*}[h]
    \centering
    \begin{subfigure}{0.49\textwidth}
        \includegraphics[width=\linewidth]{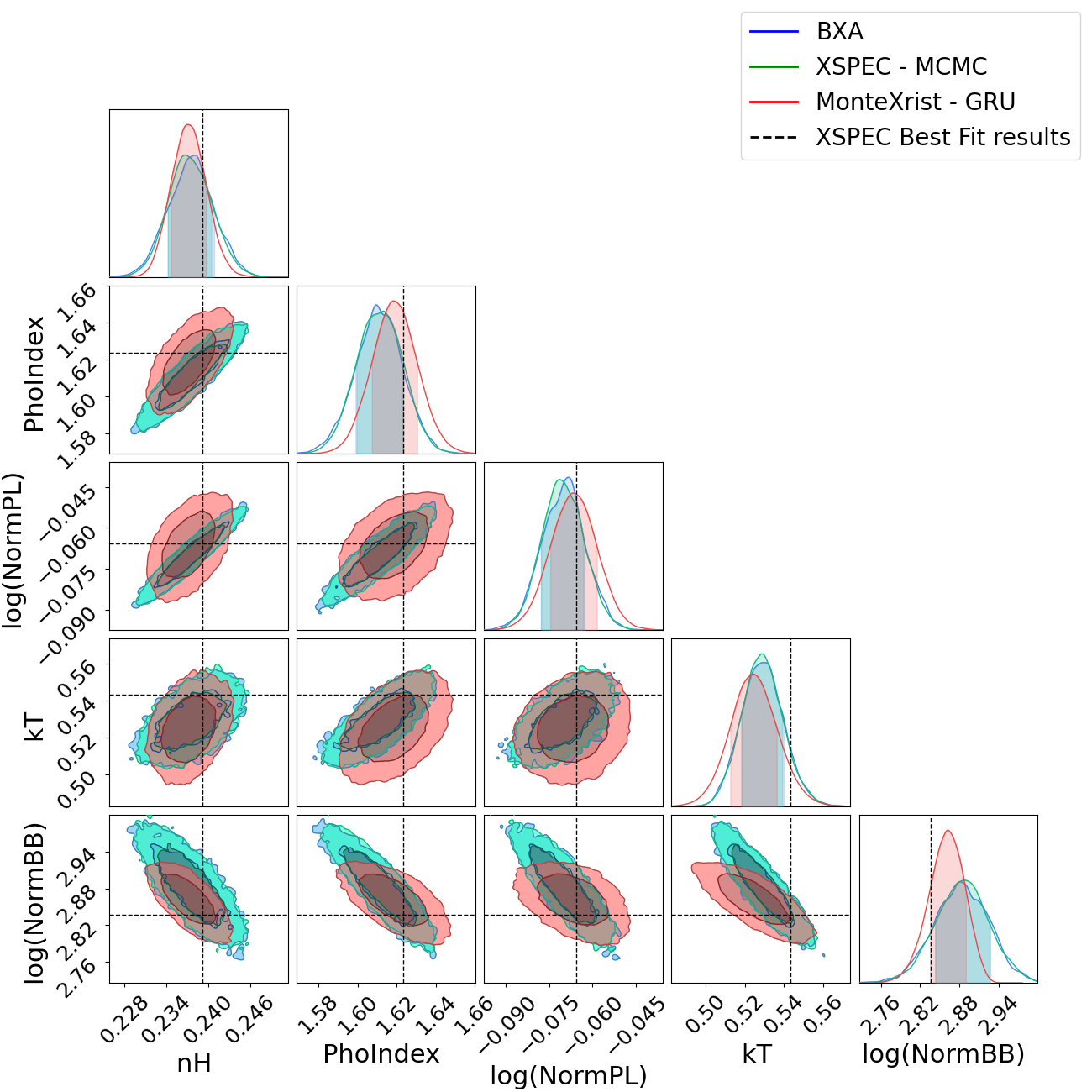}
    \end{subfigure}
    \begin{subfigure}{0.49\textwidth}
        \includegraphics[width=\linewidth]{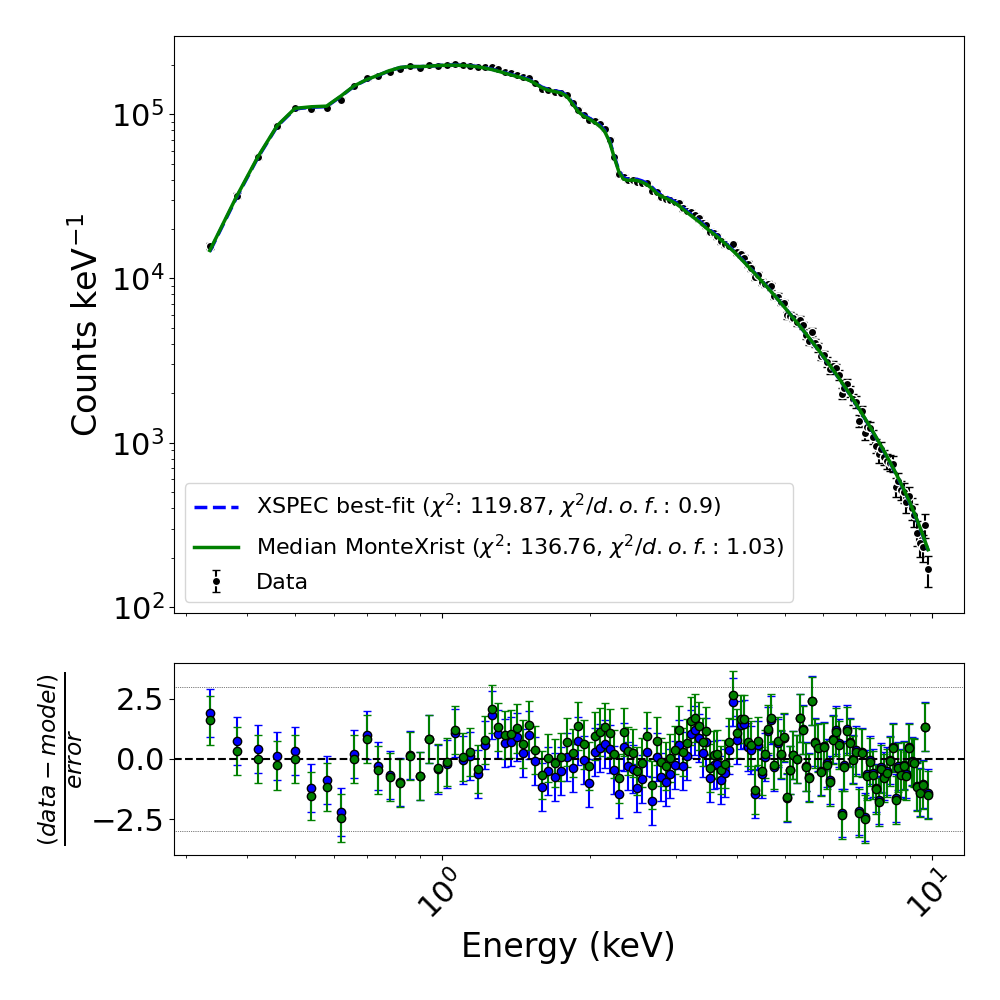}
    \end{subfigure}
    \caption{Same as Figure~\ref{fig:corner_pwl_1} and~\ref{fig:corner_pwl_10} but for the spectrum of source 4U 1820-30.}
    \label{fig:corner_bbodyrad}
\end{figure*}

In Figure~\ref{fig:corner_bbodyrad}, we present the results of our model's predictions on real data using the \textit{MonteXrist} method. The first panel (left) shows the posterior distributions of the model parameters, as inferred by our GRU-based MonteXrist approach, alongside results from \texttt{BXA} and \texttt{XSPEC MCMC}. The MonteXrist results (in red) are well-aligned with the other methods, demonstrating comparable parameter distributions and effectively capturing the correlations between parameters. The figure also highlights the model’s ability to approximate the posterior distributions typically obtained via traditional Bayesian inference approaches. Beyond achieving comparable results, a key strength of our method is its efficiency: MonteXrist, after the training phase, reaches these results in approximately one-tenth of the time required by \texttt{BXA}, using the same computational resources. This clear advantage highlights the potential of MonteXrist for faster, yet accurate, parameter estimation in X-ray astronomy.

In the second panel (right) in Figure~\ref{fig:corner_bbodyrad}, we display the best-fit emission spectrum obtained from \texttt{XSPEC}'s frequentist approach (blue dashed line) and compare it with the median prediction from \textit{MonteXrist} method (green solid line), superimposed on the observed data (black points). The lower plot shows the residuals of both fits, with \texttt{XSPEC} residuals (blue points) and MonteXrist residuals (green points) relative to the observed data.

Although both methods fit the spectral shape well, there is a slight difference in the residuals, especially at lower energies. This discrepancy arises from a higher normalization value for the blackbody component (\texttt{bbodyrad}) obtained with the MonteXrist results compared to ones obtained with the \texttt{XSPEC} fit. A higher normalization in \texttt{bbodyrad} implies a stronger contribution from the thermal component, which could account for the slightly elevated residuals observed in the MonteXrist fit at lower energies. Overall, the $\chi^2/d.o.f.$ values are $0.9$ for \texttt{XSPEC} and $1.03$ for MonteXrist, indicating that the MonteXrist approach provides results comparable to traditional methods. This illustrates that MonteXrist serves as a viable alternative for parameter estimation, providing robust uncertainty quantification through posterior distributions. The method has proven to be effective not only on simulated data but also on real observational data, highlighting its potential for practical applications in X-ray astronomy.

A possible way to enhance performance is by restricting the prior distributions of model parameters, as done in~\citet{Barret_2024}. While this approach can improve results by focusing the training on the most relevant regions of parameter space, our aim is to develop a tool with the as broad as possible applicability. By preserving full flexibility in the prior distributions, MonteXrist can be applied to a wider range of cases without requiring case-specific adjustments, making it particularly suitable for large-scale and high-resolution spectral studies.

\section{Discussion and conclusions \label{sec:discussion}}

In this work, we have developed and tested a neural network-based approach for X-ray spectral fitting, applying MCD (referred to as MonteXrist) to infer spectral model parameters and their uncertainties. We trained and validated this method on simulated data, demonstrating its ability to recover parameter distributions that are comparable to those obtained via traditional Bayesian methods, such as \texttt{MCMC} and \texttt{BXA}.
To benchmark our method, we performed tests using the same datasets as~\citet{Barret_2024}, facilitating a direct comparison between the two approaches. A possible way to refine parameter estimation is by narrowing the prior parameter space, as done in~\citet{Barret_2024} through techniques such as ResNet-based classifiers or coarse-to-fine inference. While it can improve performance by focusing the inference on the most relevant regions of parameter space, our approach is designed to operate directly on the full prior distributions. This ensures that MonteXrist remains broadly applicable, capable of handling a wide variety of datasets without requiring tailored preprocessing steps, making it particularly suitable for large-scale and high-resolution spectral studies.
This study serves as a proof of concept, illustrating the potential of neural networks for X-ray spectral analysis and setting the groundwork for future exploration. However, further testing on a broader range of models and datasets is necessary to fully assess the robustness and general applicability of this approach.

It is important to recognize the known limitations of the MCD approach. Since it approximates Bayesian inference, which may be less accurate and detailed than a full Bayesian inference, it may not capture all the nuances of the true posterior distributions. Moreover, the uncertainty estimates produced by the method are influenced by factors such as the dropout rate and the specific value the network is attempting to estimate, which may introduce additional variability in the results~\citep{Verdoja_2020}. Nonetheless, our strategy offers significant advantages for high-throughput applications, such as those expected in upcoming astronomical missions, which will produce large datasets that might require rapid and automated analysis. The MonteXrist approach mitigates the risk of local minima trapping, a common challenge in traditional spectral fitting, while achieving comparable precision to established frequentist methods without requiring the application of methods to restrict the range of the prior distribution of parameters.
Moreover, one of the key strengths of our method is its efficiency in inference time post-training. In our tests on real data, \texttt{BXA} required approximately $10$ minutes to produce a posterior chain of sufficient length, while MonteXrist completed the same task in about $1$ minute, a considerable improvement in speed. Additionally, the dataset generation process takes approximately $10$ minutes when executed on a single processor, and training the GRU model requires about one day using an NVIDIA GeForce RTX 4080. It is important to note that the code used for managing input and output operations has not been optimized for performance, leaving room for further improvements. This efficiency makes our approach particularly appealing for real-time or near-real-time applications, where rapid parameter estimation and uncertainty quantification are essential.

An interesting future extension of this work would involve the use of Principal Component Analysis (PCA)~\citep{Pearson_1901, DelSanto_2008, Jolliffe_2016} to reduce the dimensionality of the spectral data. PCA is a statistical technique that transforms the original data into a new set of uncorrelated variables, known as principal components, that capture the directions of maximum variance in the data. By retaining only the most significant components, PCA efficiently reduces the number of features while preserving most of the original data's information. In the context of spectral fitting, PCA could help decrease the input dimensionality, accelerating the training process, especially for complex architectures such as GRUs, without compromising the accuracy. This approach would be particularly beneficial for high-resolution spectra with a large number of features, where dimensionality reduction could alleviate the computational burden and improve efficiency. Previous studies, including those by \citet{Barret_2024} and \citet{Parker_2022}, have shown that PCA maintains comparable precision while significantly speeding up the training, making it a valuable tool to enhance computational performance in scenarios with high-dimensional data.

Additionally, optimizing the network architecture and hyperparameters could further enhance model performance. While we used a standard set of hyperparameters in this study, a more systematic hyperparameter tuning, especially for the dropout rate, could yield improvements in both accuracy and uncertainty quantification. Determining the optimal dropout value is crucial for balancing regularization and capturing realistic uncertainty distributions.

A compelling future direction for this research may encompass the use of Bayesian Neural Networks~\citep[BNNs;][]{MacKay_1992, Magris_2023} as an alternative to MCD for quantifying parameter uncertainties in spectral fitting tasks. Unlike the MCD approach, which approximates Bayesian inference by randomly deactivating neurons during inference, BNNs provide a true Bayesian formulation. This feature could prove particularly valuable in scenarios where accurate uncertainty estimation is critical, such as in low-count spectra or observations with high noise levels.
However, the use of BNNs also presents notable challenges. The primary drawback of BNNs is their computational expense, as methods like Variational Inference (VI) and MCMC sampling, which are commonly used to estimate the posterior distribution, can significantly increase training times. This could restrict the application of BNNs in large-scale spectral analysis or real-time data processing scenarios. Furthermore, training BNNs requires advanced Bayesian optimization techniques and often involves careful hyperparameter tuning to achieve optimal results, adding to the complexity and development time.

Despite these challenges, BNNs remain a promising alternative for uncertainty quantification in X-ray spectral fitting, especially as astronomical missions continue to produce increasingly large datasets that demand efficient, automated analysis. Future work could explore ways to integrate BNNs with high-performance computing resources and advanced optimization techniques to balance accuracy with computational feasibility. As such, BNNs offer a compelling direction for future research, with the potential to provide robust and reliable uncertainty estimates that complement the strengths of MCD in scalable, real-time applications.

In summary, our MonteXrist method demonstrates a promising alternative to traditional spectral fitting approaches, particularly for large datasets anticipated in future missions. While some challenges remain, this framework has the potential to become a valuable tool in X-ray spectral analysis, enabling fast, reliable, and interpretable inference that scales effectively with the data volume.

\begin{acknowledgements}
      We acknowledge the useful comments and suggestions of the referee, which improved the manuscript. We thank Alessandra Robba for the valuable discussions. A.T., A.P., A.Z. acknowledge supercomputing resources and support from ICSC - Centro Nazionale di Ricerca in High Performance Computing, Big Data and Quantum Computing - and hosting entity, funded by European Union - NextGenerationEU. A.T., A.N. acknowledge support from INAF Mini Grants DREAM-XS (Deep Recurrent Analysis and Modeling for X-ray Spectroscopy). C.P., F.P. acknowledge support from PRIN MUR SEAWIND funded by European Union - NextGenerationEU, INAF Grants BLOSSOM and OBIWAN (Observing high B-fIeld Whispers from Accreting Neutron stars, 1.05.23.05.12). R.L.P. acknowledges support from INAF's research grant {\it Uncovering the optical beat of the fastest magnetised neutron stars (FANS)} and from the Italian Ministry of University and Research (MUR), PRIN 2020 (prot. 2020BRP57Z) \textit{Gravitational and Electromagnetic-wave Sources in the Universe with current and next-generation detectors (GEMS)}. E.A. acknowledges funding from the Italian Space Agency, contract ASI/INAF no. I/004/11/4.
\end{acknowledgements}


\bibliography{biblio}
\bibliographystyle{aa}

\end{document}